\newif\iftwocolumnversion
\twocolumnversiontrue

\iftwocolumnversion
  \documentclass[aps,prapplied,twocolumn,superscriptaddress]{revtex4-2}
\else
  \documentclass[aps,prapplied,preprint,superscriptaddress,linenumbers]{revtex4-2}
\fi

\usepackage{bm}
\usepackage{amsmath}
\usepackage{amssymb}
\usepackage{braket}
\usepackage{makecell}
\usepackage{tikz}
\usetikzlibrary{quantikz}
\usetikzlibrary{decorations.pathreplacing}
\usepackage{mathdots}
\usepackage{mathtools}
\usepackage{graphicx}
\usepackage{longtable}
\usepackage[colorlinks=true, linkcolor=blue, citecolor=blue, urlcolor=blue]{hyperref}

\newcommand{\twocolbreak}{\iftwocolumnversion\expandafter\\\fi}

\usepackage[final]{changes}

\begin{document}

\title{Quantum algorithm for the collision-coalescence of cloud droplets}

\author{Kazumasa Ueno}
\email{kazumasa-e67@eps.s.u-tokyo.ac.jp}
\affiliation{The University of Tokyo}
\author{Hiroaki Miura}
\affiliation{The University of Tokyo}

\date{\today}

\begin{abstract}
\iftwocolumnversion\else\linenumbers\fi
Quantum computing may help reduce computational costs of simulating large and nonlinear systems, but research into the use of quantum computers in atmospheric and oceanic sciences is still at an early stage.
This study explores the use of quantum computing for calculating the collision-coalescence process that dominates the size growth of liquid particles in cloud microphysics.
Inspired by the quantum algorithms developed in the field of financial engineering, we propose a new algorithm based on a master equation that describes the time evolution of the droplet mass distribution.
Our algorithm uses the quantum amplitudes to encode the probability distribution of droplet mass and calculates the expected number of droplets via the quantum amplitude estimation. The key contribution is an encoding strategy that maps the multivariate collision-coalescence problem onto a quantum computation by recording only the transition history rather than the full mass distributions at every time step. This approach reduces the per-step qubit cost from $\mathcal{O}(\sqrt{N})$ to $\mathcal{O}(\log N)$, where $N$ is the number of mass bins.
Our resource analysis shows that the number of T gates (T-count) scales as $\widetilde{\mathcal{O}}(N^2)$ per mass bin in $N$, for a fixed number of time steps and a given target accuracy, where the $\widetilde{\mathcal{O}}$ notation suppresses polylogarithmic factors in $N$.
This is a substantial improvement over classical master-equation methods, whose cost grows super-polynomially with $N$ as the number of possible states increases.
Our results suggest that the collision-coalescence process is one of the promising targets of quantum computing in the field of atmospheric science.
\end{abstract}


\maketitle

\section{Introduction}
Numerical modeling is an essential approach for understanding the atmosphere-ocean system.
This is because the system is too complex to be fully understood through analytical methods, and direct experiments of the system are impossible.
Our understanding of atmospheric and oceanic phenomena has been advancing alongside the development of computers.
For example, advances in computing have enabled climate-model projections to become more complex, with more detailed representations of physical processes at higher resolution.
Despite the inherent limitations of numerical models, these models can compute responses of the climate system to the rapid increase of the greenhouse-gas concentrations, and provide useful insights into future climate change~\cite{Randall2001-vy,Ghil2020-gn,IPCC2023-1,IPCC2023-2,IPCC2023-3}.

It is important to handle stochasticity in those simulations to enhance their realism.
This is because the system is nonlinear and chaotic, and thus, even small perturbations can grow into large differences in the future.
Stochasticity can arise from several sources and can be represented in diverse manners.
For example, ensemble simulations are widely used to account for uncertainties arising from initial conditions and intrinsic fluctuations~\cite{Parker2010-ub,Mizuta2017-iz,Tegegne2020-sf}.
Stochastic representations are also used in parameterizations that incorporate effective interactions between grid-scale (resolved) processes and subgrid-scale (unresolved) processes in numerical models.
Incorporating stochastic effects into parameterizations improves representations of the resolved flow, enhances the accuracy of the mean-state and noise-induced regime transitions, and enables a more realistic response to external forcing~\cite{Berner2017-qc}.
There are various approaches to incorporating stochastic effects; the super-parameterization method~\cite{Grabowski1999-oy,Grabowski2004-nq}, stochastic parameterizations~\cite{Buizza1999-ae,Palmer2009-yo,Qiao2017-za}, and methods based on probability distributions~\cite{Khain2004-ro,Wang2006-tz,Morrison2010-jy}.
The super-parameterization represents subgrid-scale processes by embedding a reduced-dimension higher-resolution model within the main model.
Stochastic parameterizations introduce stochastic perturbations to account for the influence of the unresolved variability.
Methods based on probability distributions explicitly consider the probability distributions of the quantities of interest. This third approach has been primarily applied in cloud microphysics.

Cloud microphysics describes the phase change of water in the atmosphere and is among the most important processes in atmospheric models.
It represents the growth of cloud droplets inside clouds by the collision-coalescence process. 
This process is associated with the cloud properties such as the droplet size distribution, the droplet number concentration, and the optical properties.
Accurate representations of microphysical processes, however, remain a significant challenge.
One of the key difficulties is the inherent stochasticity of cloud microphysics, and another is the wide range of sizes that cloud droplets span~\cite{Grabowski2019-nx}.

In cloud microphysics, there are three approaches to represent its stochasticity: the spectral bin method~\cite{Khain2004-ro,Wang2006-tz,Morrison2010-jy}, the super-droplet method~\cite{Shima2009-ze,Shima2020-lw}, and the master equation method~\cite{Gillespie1972-gk,Alfonso2015-vq}.
Each of these methods is designed to capture different aspects of cloud droplet evolution, and has its strengths and limitations due to its design.
The spectral bin method discretizes droplet size into bins to predict the time evolution of the droplet size distribution.
In this method, the stochasticity is incorporated through the ensemble-mean size distribution, and fluctuations of the distribution are not explicitly represented.
The super-droplet method provides a more complete representation by partially accounting for fluctuations in the droplet size distribution.
The master equation method considers the probability distribution of each realizable droplet size distribution.
While the master equation method is the closest to the first principles among the three methods, solving the master equation is computationally challenging because the number of possible droplet size distributions grows super-polynomially with the number of bins.
The Monte Carlo method has been used to solve the master equation~\cite{Landau2014-vs}, but it requires a large number of calculations.

In the area of financial engineering, various quantum algorithms have been proposed to solve stochastic differential equations on fault-tolerant quantum computers~\cite{Rebentrost2018-xu,Herman2023-lc}.
These algorithms are used to calculate the time evolution of financial derivatives and to determine the optimal options for maximizing returns.
The underlying assumption in financial engineering is that the variations of derivatives follow master equations.
The quantum computation can efficiently handle higher-dimensional probability distributions by utilizing the properties of quantum dynamics such as the quantum superposition, interference, and entanglement.
In addition, the quantum amplitude estimation~\cite{Brassard-ut,Grinko2021-nk} that enables efficient computations of key characteristics from probability distributions further enhances the applicability of the quantum algorithms in financial modeling.

Our work here is inspired by quantum computing in financial modeling~\cite{Wang2024-me}. We develop a quantum algorithm to solve the master equation that governs the collision-coalescence process of cloud droplets.
We can utilize the same principles as those used in quantum-enhanced financial computations for handling higher-dimensional probability distributions.

The existing quantum algorithms in financial modeling compute the time evolution of a probability distribution but typically focus on a single stochastic variable (e.g., an asset price).
In contrast, cloud microphysics requires a multivariate description, because the droplet numbers in all $N$ mass bins evolve jointly.
This work makes two main contributions.
The primary contribution is a problem formulation and encoding strategy: we recast the classical collision-coalescence master equation into a quantum-computable form by encoding the probability distribution in quantum amplitudes and, crucially, recording only the transition history rather than the complete set of mass distributions at every time step.
While a naive encoding that tracks all possible mass distributions would require $\mathcal{O}(\sqrt{N})$ qubits per time step, our transition-history encoding requires only $\mathcal{O}(\log N)$ qubits per time step.
Our second contribution is the quantum algorithm itself and is more incremental: we combine the fixed-point quantum arithmetic~\cite{Haner2018-jy} and the quantum amplitude estimation~\cite{Brassard-ut,Grinko2021-nk} in a form tailored to the collision-coalescence problem.
The novelty lies in the formulation and encoding of the collision-coalescence master equation, and in the circuit construction that redistributes the probability over all possible states at once, whereas a classical computation must proceed state by state.
This efficient handling of multivariate distributions is not only relevant to cloud droplet evolution but also to other stochastic systems with complex interactions such as chemical reaction networks.

This paper is organized as follows.
Section~\ref{Formulation} summarizes the formulation of the cloud collision-coalescence process based on the master equation in~\cite{Alfonso2015-vq}.
Section~\ref{Quantum algorithm} describes the new quantum algorithm to solve the master equation. Section~\ref{Quantum resource estimates} estimates the quantum resources required for the new algorithm. We focus on the T-count and the number of logical qubits, and analyze how they scale with the problem size (e.g., the number of mass bins and time steps).
Sections~\ref{Discussion} and~\ref{Conclusion} discuss implications of our quantum algorithm for cloud microphysics and conclude the study.

\section{\label{Formulation}Formulation}

Cloud droplets span a wide range of sizes, from micrometers to millimeters in radius.
Their size distribution evolves through the collision-coalescence~\cite{Rogers1996-ls}, in which droplets stochastically collide and merge to form larger droplets.
Several approaches have been developed to describe collision-coalescence, including the stochastic collection equation (SCE), the particle-based methods such as the super-droplet method (SDM), and the master equation method~\cite{Morrison2020-av}.
In this study, we adopt the master equation method as a fundamental probabilistic description of the collision-coalescence process.

Here, we briefly compare the strengths and limitations of the SCE and the SDM.
The SCE evolves the ensemble-mean droplet size distribution. It is widely used because it provides a computationally efficient description of collision-coalescence~\cite{Khain2015-tw}.
However, it can predict only the mean behavior and cannot capture realization-to-realization variability.
In a finite cloud volume, droplet collisions occur as discrete random events, so the size distribution can vary substantially from one realization to another.
In particular, rare sequences of collisions can produce unusually large droplets early in the process and accelerate the growth of the distribution tail~\cite{Dziekan2017-wb}.
In contrast, the SDM represents collisions stochastically and can capture realization-to-realization variability~\cite{Shima2009-ze}.
However, a large droplet population is represented using a finite number of super-droplets with multiplicities, and collisions are realized by stochastic sampling.
As a result, statistics of the large-droplet tail are sensitive to the number of super-droplets, because sampling noise decreases only as the particle number increases~\cite{Dziekan2017-wb}.
Therefore, neither the deterministic mean-field approach (i.e., SCE) nor the sampling-based particle approach (i.e., SDM) can directly evolve the full probability distribution of droplet populations.

In contrast, the master equation method directly evolves the full probability distribution over droplet populations.
Although the solution of the master equation is computationally challenging for large systems, it is useful because it can provide an explicit reference for the evolution of probabilities to assess the validity of approximations inherent in SCE and SDM~\cite{Alfonso2015-vq,Dziekan2017-wb}.
This study tackles this computational challenge using quantum computing to accelerate the solution of the master equation.
The remainder of this section summarizes the master-equation formulation of collision-coalescence, following previous studies~\cite{Gillespie1972-gk,Alfonso2015-vq}.

We represent the cloud droplet mass distribution discretely by grouping droplets according to mass.
We assume the existence of the fundamental unit mass, $x_1$, and consider that each droplet mass is an integer multiple of it. This unit mass is the physically meaningful minimal mass scale for discretization.
Then, the mass of the cloud droplet in the $i$-th bin is given by $x_i = i x_1$ $(i=1,2,\ldots,N)$, where $N$ is the number of mass bins.
If $n_i$ denotes the number of droplets in the $i$-th bin, the vector $\bar{\bm{n}} = (n_1, n_2, \ldots, n_N)$ fully describes the state of the system.
As a simplifying assumption, we assume that the total droplet mass $x_\text{total}$ is conserved throughout the time evolution.
This assumption allows us to focus on the collision-coalescence process by neglecting mass changes caused by condensation and evaporation.
We also neglect breakups of droplets for simplicity, so that the evolution is governed solely by coalescence.
Here, we choose $N$ such that $x_\text{total} \leq x_N$.
Consequently, droplets with masses greater than $x_N$ cannot occur, and thus, $n_j=0\ (j>N)$.

We denote the probability of a given state as $P(\bar{\bm{n}})$. Then the master equation governing the time evolution of the state is given by~\cite{Tanaka1993-ft}:
\begin{equation}
  \iftwocolumnversion
  \begin{aligned}
    &\frac{\partial P(\bar{\bm{n}};t)}{\partial t} \\
    &\quad= \sum_{i=1}^N\sum_{j=i+1}^{N} K(i,j)(n_i+1)(n_j+1)
    \\ &\qquad\qquad
    P(\ldots,n_i+1,\ldots,n_j+1,\ldots,n_{i+j}-1,\ldots;t) \\
    &\qquad+ \sum_{i=1}^N\frac{1}{2}K(i,i)(n_i+2)(n_i+1)
    \\ &\qquad\qquad
    P(\ldots,n_i+2,\ldots,n_{2i}-1,\ldots;t) \\
    &\qquad-\sum_{i=1}^{N}\sum_{j=i+1}^N K(i,j)n_in_jP(\bar{\bm{n}};t)
    \\&\qquad
    -\sum_{i=1}^N\frac{1}{2}K(i,i)n_i(n_i-1)P(\bar{\bm{n}};t).
  \end{aligned}
  \else
  \begin{aligned}
    \frac{\partial P(\bar{\bm{n}};t)}{\partial t}
    &= \sum_{i=1}^N\sum_{j=i+1}^{N} K(i,j)(n_i+1)(n_j+1)
    P(\ldots,n_i+1,\ldots,n_j+1,\ldots,n_{i+j}-1,\ldots;t) \\
    &\qquad+ \sum_{i=1}^N\frac{1}{2}K(i,i)(n_i+2)(n_i+1)
    P(\ldots,n_i+2,\ldots,n_{2i}-1,\ldots;t) \\
    &\qquad-\sum_{i=1}^{N}\sum_{j=i+1}^N K(i,j)n_in_jP(\bar{\bm{n}};t)
    -\sum_{i=1}^N\frac{1}{2}K(i,i)n_i(n_i-1)P(\bar{\bm{n}};t).
  \end{aligned}
  \fi
  \label{eq:master}
\end{equation}
This equation describes how the rate of change in the probability of a state is determined by the probabilities of the associated states.
The first term on the right-hand-side represents the probability gain due to the collision between droplets in the $i$-th and $j$-th bins, which results in an increase in the $(i+j)$-th droplet by one.
The second term accounts for the probability gain caused by the collision of two droplets in the $i$-th bin.
The last two terms represent the loss of probability from the target state $P(\bar{\bm{n}})$ due to droplet collisions.

The coefficients in the form of $K(i,j)n_in_j$ in each term represent transition probabilities. $K(i,j)$ is the collection kernel that represents the probability of collisions between the two droplets of masses $x_i$ and $x_j$.
These coefficients can be separated into two components: the droplet-number-dependent part, $n_in_j$, and the droplet-number-independent part, $K(i,j)$.
This separation is essentially important for our quantum algorithm and its details will be described later.

Solving Eq.~\eqref{eq:master} for all possible realizations of $\bar{\bm{n}}$ is computationally demanding.
This is because the number of possible states, $R(N)$, grows rapidly with $N$ as follows~\cite{Hall2011-sw}: 
\begin{equation}
  \label{eq:state_number}
  R(N) \sim \frac{1}{4N\sqrt{3}}\exp{\left(\pi\left(\frac{2N}{3}\right)^{1/2}\right)}.
\end{equation} 
Although the super-polynomial growth in $R(N)$ has made it infeasible to perform this calculation for $N>40$ on classical computers~\cite{Alfonso2017-fg}, quantum computers might make it feasible.

In our calculation, the time derivative in Eq.~\eqref{eq:master} is discretized using a first-order finite difference scheme, following ~\cite{Alfonso2015-vq}.
Then, the discrete form of the master equation becomes
\begin{equation}
  \iftwocolumnversion
  \begin{aligned}
    P&(\bar{\bm{n}};t_0+\Delta t) \\
    &= P(\bar{\bm{n}};t_0) \\
    &\qquad+ \Delta t \sum_{i=1}^N\sum_{j=i+1}^{N} K(i,j)(n_i+1)(n_j+1)
    \\ &\qquad\qquad
    P(\ldots,n_i+1,\ldots,n_j+1,\ldots,n_{i+j}-1,\ldots;t_0) \\
    &\qquad+ \Delta t \sum_{i=1}^N\frac{1}{2}K(i,i)(n_i+2)(n_i+1)
    \\ &\qquad\qquad
    P(\ldots,n_i+2,\ldots,n_{2i}-1,\ldots;t_0) \\
    &\qquad- \Delta t \sum_{i=1}^{N}\sum_{j=i+1}^N K(i,j)n_in_jP(\bar{\bm{n}};t_0)
    \\&\qquad
    - \Delta t \sum_{i=1}^N\frac{1}{2}K(i,i)n_i(n_i-1)P(\bar{\bm{n}};t_0).
  \end{aligned}
  \else
  \begin{aligned}
    P&(\bar{\bm{n}};t_0+\Delta t) \\
    &= P(\bar{\bm{n}};t_0) \\
    &\qquad+ \Delta t \sum_{i=1}^N\sum_{j=i+1}^{N} K(i,j)(n_i+1)(n_j+1)
    P(\ldots,n_i+1,\ldots,n_j+1,\ldots,n_{i+j}-1,\ldots;t_0) \\
    &\qquad+ \Delta t \sum_{i=1}^N\frac{1}{2}K(i,i)(n_i+2)(n_i+1)
    P(\ldots,n_i+2,\ldots,n_{2i}-1,\ldots;t_0) \\
    &\qquad- \Delta t \sum_{i=1}^{N}\sum_{j=i+1}^N K(i,j)n_in_jP(\bar{\bm{n}};t_0)
    - \Delta t \sum_{i=1}^N\frac{1}{2}K(i,i)n_i(n_i-1)P(\bar{\bm{n}};t_0).
  \end{aligned}
  \fi
  \label{eq:master_discretized}
\end{equation}
The system is advanced in time using Eq.~\eqref{eq:master_discretized}.
Repeating this procedure yields the probability distribution over all possible states at the final time $t_\text{final}$.

Usually, quantities of our interest are not the probabilities themselves, but rather characteristic values derived from the probabilities.
For example, the probability of obtaining $n_i=n$ at time $t$ is expressed as 
\iftwocolumnversion
\begin{multline}
  \label{eq:probability}
  P(n_i=n;t) \\
  = \sum_{\{n_j\}_{j \neq i}} P(n_1, \ldots, n_{i-1}, n_i=n, n_{i+1}, \ldots,n_N;t),
\end{multline}
\else
\begin{equation}
  \label{eq:probability}
  P(n_i=n;t) = \sum_{\{n_j\}_{j \neq i}} P(n_1, \ldots, n_{i-1}, n_i=n, n_{i+1}, \ldots,n_N;t),
\end{equation}
\fi
where the summation is taken over all $n_j$ with $j \in \{1, \ldots, N\} \setminus \{i\}$.
Similarly, the expected number of droplets in the $i$-th bin at time $t$ can be calculated using Eq.~\eqref{eq:probability} as 
\begin{equation}
  \label{eq:expected_value}
  \braket{n_i} = \sum_n nP(n_i=n;t).
\end{equation}
A key advantage of our quantum algorithm is its ability to compute these characteristic values efficiently.

\section{\label{Quantum algorithm}Quantum algorithm}
\begin{figure*}[t]
  \centering
  \includegraphics[width=\textwidth]{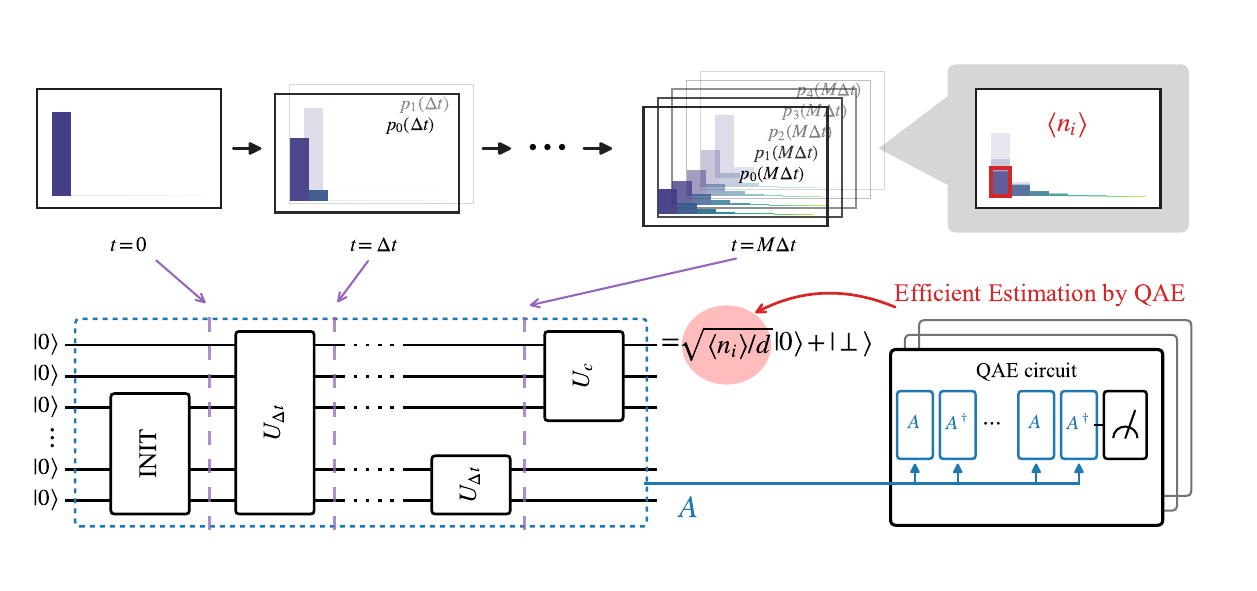}
  \caption{Overview of the proposed quantum algorithm for the collision-coalescence process.
  Top: time evolution of the droplet mass distribution. At $t=0$, all droplets reside in the first bin. At each subsequent time $t=\Delta t,\ldots,M\Delta t$, the state is a probability-weighted superposition of mass distributions. Here, $M$ is the number of time steps, and $p_k(t)$ is the probability weight of the $k$-th mass distribution.
  Bottom: the corresponding quantum circuit. The operation $A$ consists of INIT, $M$ applications of $U_{\Delta t}$, and $U_c$.
  The state of the auxiliary qubit is written as $\sqrt{\braket{n_i}/d}\,\ket{0}+\ket{\perp}$, where $\braket{n_i}$ is the expected number of droplets in the $i$-th bin. The normalization factor is $d=2^{q_i}$, where $q_i$ is the number of qubits of the register $\mathcal{N}_i$ that stores the droplet number of the $i$-th bin. The term $\ket{\perp}$ represents the component orthogonal to $\ket{0}$ on the auxiliary qubit.
  The meter symbol indicates the auxiliary-qubit measurement used in quantum amplitude estimation (QAE).}
  \label{fig:concept}
\end{figure*}

We first present an overview of the proposed algorithm in Fig.~\ref{fig:concept}. The operator INIT prepares the initial droplet mass distribution in the quantum registers. The operator $U_{\Delta t}$ propagates the probability distribution by one time step and is applied $M$ times, where $M$ is the number of time steps, producing a superposition of the reachable mass distributions, with amplitudes given by the square roots of their probabilities. The operator $U_c$ then encodes the quantity of interest in the amplitude of an auxiliary qubit. In this example, the quantity of interest is the expected droplet number $\braket{n_i}$. These operations together define a unitary operator $A$. Quantum amplitude estimation (QAE) uses $A$ and its inverse $A^\dagger$ to estimate the encoded amplitude to the desired accuracy. The individual operations are described below.

\subsection{Quantum state representation}
In our algorithm, the probability distribution over droplet mass distributions is encoded in the quantum state $\ket{\psi(t)}_{\mathcal{H}\mathcal{N}}$ as
\begin{equation}
  \label{eq:quantum_state}
  \ket{\psi(t)}_{\mathcal{H}\mathcal{N}} = \sum_{\bm{h}}\sum_{\bar{\bm{n}}} \sqrt{P(\bm{h},\bar{\bm{n}};t)}\ket{\bm{h}}_{\mathcal{H}}\ket{\bar{\bm{n}}}_{\mathcal{N}},
\end{equation}
where the vector $\ket{\bm{h}}_{\mathcal{H}}$ represents a transition history and the vector $\ket{\bar{\bm{n}}}_{\mathcal{N}}$ represents a mass distribution.
The subscripts $\mathcal{H}$ and $\mathcal{N}$ indicate the quantum registers storing these components.
The transition history $\ket{\bm{h}}_{\mathcal{H}}$ contains details of the sequence of transitions and is expressed as
\begin{equation}
  \label{eq:transition_history}
  \ket{\bm{h}}_{\mathcal{H}} = \ket{h_1,h_2,\ldots,h_M}_{\mathcal{H}} = \ket{h_1}_{\mathcal{H}_1}\ket{h_2}_{\mathcal{H}_2}\ldots \ket{h_M}_{\mathcal{H}_M},
\end{equation}
where $M$ is the number of time steps, $h_m$ is a transition label at the $m$-th time step that uniquely specifies the pair of bins involved in the droplet collisions, and the subscript $m$ is an integer index satisfying $1 \le m \le M$.
The qubits $\mathcal{H}_m$ constitute a quantum register to store the $m$-th transition history.
Generally, $\mathcal{H}_m$ does not consist of a single qubit, but the number of qubits used for $\mathcal{H}_m$ is approximately equal to the logarithm of the total number of transition labels.
The details of this count will be described in Sec.~\ref{Quantum resource estimates}.
Similarly, the mass distribution $\ket{\bar{\bm{n}}}_{\mathcal{N}}$ is given by
\begin{equation}
  \label{eq:mass_distribution}
  \ket{\bar{\bm{n}}}_{\mathcal{N}} = \ket{n_1,n_2,\ldots,n_N}_{\mathcal{N}} = \ket{n_1}_{\mathcal{N}_1}\ket{n_2}_{\mathcal{N}_2}\ldots \ket{n_N}_{\mathcal{N}_N}.
\end{equation}
Let $P(\bm{h},\bar{\bm{n}};t)$ indicate the probability for the system to be in a combined state denoted by $\bm{h}$ and $\bar{\bm{n}}$ at time $t$.
Taking the sum of $P(\bm{h},\bar{\bm{n}};t)$ with respect to $\bm{h}$, we obtain the following relation: 
\begin{equation}
  P(\bar{\bm{n}};t) = \sum_{\bm{h}}P(\bm{h},\bar{\bm{n}};t).
\end{equation}

In this study, we consider a closed system in which the total mass $x_{\text{total}}=x_N$ is conserved, and we assume an initial condition where all droplets reside in the first bin.
Thus, $n_1=x_{\text{total}}/x_1=N$, and the probability of the initial state is $P(n_1=N,n_2=0,\ldots,n_N=0;t=0)=1$.
The corresponding quantum state is given by
\begin{equation}
  \label{eq:initial_state}
  \ket{\psi(0)}_{\mathcal{HN}} = \ket{\bm{0}}_{\mathcal{H}}\ket{N}_{\mathcal{N}_1}\ket{0}_{\mathcal{N}_2}\ldots\ket{0}_{\mathcal{N}_N},
\end{equation}
where $\ket{\bm{0}}_{\mathcal{H}} = \ket{0}_{\mathcal{H}_1}\ket{0}_{\mathcal{H}_2}\ldots\ket{0}_{\mathcal{H}_M}$ represents the initial transition history with no transitions. 

\subsection{Probability division}
The time evolution of the probability distribution of cloud droplet masses can be obtained by tracking the evolving quantum state.
At each time step, the probabilities are redistributed according to the transition labels.
We call this redistribution process for all possible states as the \textit{probability division} in this study.
For example, after a single time step $\Delta t$ from the initial state, the quantum state given by Eq.~\eqref{eq:initial_state} evolves as
\begin{equation}
  \iftwocolumnversion
  \begin{aligned}
    &\ket{\psi(\Delta t)}_{\mathcal{HN}}
    \\
    &= \sqrt{1-\frac{1}{2}K(1,1)N(N-1)\Delta t}
    \\&\qquad
    \ket{0}_{\mathcal{H}_1}\ket{0}_{\mathcal{H}_2}\ldots\ket{0}_{\mathcal{H}_M}\ket{N}_{\mathcal{N}_1}\ket{0}_{\mathcal{N}_2}\ldots\ket{0}_{\mathcal{N}_N} \\
    &\quad+ \sqrt{\frac{1}{2}K(1,1)N(N-1)\Delta t}
    \\&\qquad
    \ket{1}_{\mathcal{H}_1}\ket{0}_{\mathcal{H}_2}\ldots\ket{0}_{\mathcal{H}_M}\ket{N-2}_{\mathcal{N}_1}\ket{1}_{\mathcal{N}_2}\ldots\ket{0}_{\mathcal{N}_N}.
  \end{aligned}
  \else
  \begin{aligned}
    \ket{\psi(\Delta t)}_{\mathcal{HN}}
    &= \sqrt{1-\frac{1}{2}K(1,1)N(N-1)\Delta t}
    \ket{0}_{\mathcal{H}_1}\ket{0}_{\mathcal{H}_2}\ldots\ket{0}_{\mathcal{H}_M}\ket{N}_{\mathcal{N}_1}\ket{0}_{\mathcal{N}_2}\ldots\ket{0}_{\mathcal{N}_N} \\
    &\quad+ \sqrt{\frac{1}{2}K(1,1)N(N-1)\Delta t}
    \ket{1}_{\mathcal{H}_1}\ket{0}_{\mathcal{H}_2}\ldots\ket{0}_{\mathcal{H}_M}\ket{N-2}_{\mathcal{N}_1}\ket{1}_{\mathcal{N}_2}\ldots\ket{0}_{\mathcal{N}_N}.
  \end{aligned}
  \fi
\end{equation}
In this first step, we need to consider only the collision of two droplets in the first bin.
We label this transition as $1$.
We reserve the label $0$ for the no-transition case.
The numbering rule of the transition labels will be explained later.

Figure~\ref{fig:probability_division} schematically shows the sequence of the probability divisions for multiple steps starting from the initial state.
This includes both the collision transitions (with labels $h \geq 1$) and the non-collision holding events ($h=0$).

We can expect a quantum advantage in this probability division.
In the classical algorithm, the probability division is performed sequentially for all possible states.
The number of possible states grows super-polynomially, as given by Eq.~\eqref{eq:state_number}, so the computational cost grows accordingly because of this sequential execution.
In contrast, quantum computing can process these state transitions simultaneously.
By exploiting the quantum superposition, a single probability division applies to all possible states at once, and the repeated application of the probability division is unnecessary.
This is one of the key advantages of our quantum algorithm.
The parallel handling of all states significantly reduces computational costs.

\begin{figure*}
  \centering
  \resizebox{\textwidth}{!}{
    \begin{tikzpicture}
      \node[text width=3cm, align=center] at (0, 2) {$(N,0,\ldots,0)$};
      \node[text width=3cm, align=center] at (5, 3) {$(N,0,\ldots,0)$};
      \node[text width=3cm, align=center] at (5, 1) {$(N-2,1,\ldots,0)$};
      \node[text width=3cm, align=center] at (11, 4) {$(N,0,\ldots,0)$};
      \node[text width=3cm, align=center] at (11, 3) {$(N-2,1,\ldots,0)$};
      \node[text width=3cm, align=center] at (11, 2) {$(N-2,1,\ldots,0)$};
      \node[text width=3cm, align=center] at (11, 1) {$(N-4,2,\ldots,0)$};
      \node[text width=3.5cm, align=center] at (11, 0) {$(N-3,0,1,\ldots,0)$};
      
      \draw[-, thick] (1.2,2) -- (1.5,2);
      \draw[-, thick] (1.5,3) -- (1.5,1);
      \draw[->, thick] (1.5,3) -- (3.5,3);
      \draw[->, thick] (1.5,1) -- (3.5,1);
      \draw[-, thick] (7,4) -- (7,3);
      \draw[-, thick] (7,0) -- (7,2);
      \draw[->, thick] (6.5,3) -- (9,3);
      \draw[->, thick] (6.5,1) -- (9,1);
      \draw[->, thick] (7,4) -- (9,4);
      \draw[->, thick] (7,2) -- (9,2);
      \draw[->, thick] (7,0) -- (9,0);
      
      \node at (2.5,3.3) {$r_0(\bar{\bm{n}}_0)$};
      \node at (2.5,1.3) {$r_1(\bar{\bm{n}}_0)$};
      \node at (8,4.3) {$r_0(\bar{\bm{n}}_0)$};
      \node at (8,3.3) {$r_1(\bar{\bm{n}}_0)$};
      \node at (8,2.3) {$r_0(\bar{\bm{n}}_1)$};
      \node at (8,1.3) {$r_1(\bar{\bm{n}}_1)$};
      \node at (8,0.3) {$r_2(\bar{\bm{n}}_1)$};
      
      \node at (0,2.5) {$\bar{\bm{n}}_0$};
      \node at (5,1.5) {$\bar{\bm{n}}_1$};
      
      \node at (13,4) {$\cdots$};
      \node at (13,3) {$\cdots$};
      \node at (13,2) {$\cdots$};
      \node at (13,1) {$\cdots$};
      \node at (13,0) {$\cdots$};
      
      \node at (0,-1) {$t=0$};
      \node at (5,-1) {$t=\Delta t$};
      \node at (11,-1) {$t=2\Delta t$};
    \end{tikzpicture}
    }
    \caption{The schematic diagram of the probability-division process over multiple steps.
    The numbers in parentheses represent the mass distribution $\bar{\bm{n}}$.
    For convenience, the states $(N,0,\ldots,0)$ are denoted as $\bar{\bm{n}}_0$, and $(N-2,1,\ldots,0)$ as $\bar{\bm{n}}_1$.
    The arrows indicate the probability flow.
    The transition probability from the state $\bar{\bm{n}}_i$ by the transition with label $h$ is denoted as $r_h(\bar{\bm{n}}_i)$ above the arrows.
    The probability of no transition is represented by $r_0$. The transition probabilities of the transitions with labels $1$ and $2$ are denoted as $r_1$ and $r_2$, respectively.}
    \label{fig:probability_division}
  \end{figure*}
  
We next describe how a single time step is implemented using the probability-division procedure.
To advance the calculation by one time step, probability division is applied to every possible transition from every possible state.
For example, the probability flow from a state $(n_1=N-2,n_2=1,n_3=0,\ldots)$ to a state $(n_1=N-3,n_2=0,n_3=1,\ldots)$ and from a state $(n_1=N-4,n_2=2,n_3=0,\ldots)$ to a state $(n_1=N-5,n_2=1,n_3=1,\ldots)$ are conducted simultaneously because both are associated with collisions between droplets in the first and the second bins to create droplets in the third bin.
Thus, probability division can be performed simultaneously for all states sharing the same transition label.

\begin{figure*}
  \resizebox{\textwidth}{!}{
  \begin{tikzpicture}
    \node[text width=2cm, align=center] (A) at (0, 2) {$P(\bar{\bm{n}})$};
    \node[text width=2cm, align=center] (B) at (3, 2) {$s_{H}P(\bar{\bm{n}})$};
    \node[text width=2cm, align=center] (C) at (6, 2) {$s_{H-1}P(\bar{\bm{n}})$};
    \node[text width=1cm, align=center] (D) at (8.5, 2) {$\cdots$};
    \node[text width=2cm, align=center] (E) at (11, 1.75) {$\underbrace{s_1}_{r_0}P(\bar{\bm{n}})$};

    \node[text width=2cm, align=center] (F) at (1.5, 0.5) {$r_HP(\bar{\bm{n}})$};
    \node[text width=2.5cm, align=center] (G) at (4.5, 0.25) {$\underbrace{r'_{H-1}s_H}_{\smash{r_{H-1}}}P(\bar{\bm{n}})$};
    \node[text width=2cm, align=center] (H) at (8.5, 0.5) {$\cdots$};

    \draw[->, thick] (1,2) -- (2,2);
    \draw[->, thick] (4,2) -- (5,2);
    \draw[->, thick] (7,2) -- (8,2);
    \draw[->, thick] (9,2) -- (10,2);

    \draw[->, thick] (1.5,2) -- (1.5,1);
    \draw[->, thick] (4.5,2) -- (4.5,1);
    \draw[->, thick] (7.5,2) -- (7.5,1);
    \draw[->, thick] (9.5,2) -- (9.5,1);

    \draw[thick,dashed,draw=blue]
      (0.1, -0.4) -- (11.9, -0.4) -- (11.9, 2.5) -- (10.1, 2.5) -- (10.1, 0.9) -- (0.1, 0.9) -- cycle;
    
    \node[color=blue] at (8.5,-0.9) {The probability after a time interval $\Delta t$};
  \end{tikzpicture}%
  }%
  \caption{The schematic diagram of the probability-division process for a single time step.
  The probability $P(\bar{\bm{n}})$ is first divided into $s_H P(\bar{\bm{n}})$ and $r_H P(\bar{\bm{n}})$, where $H$ is the total number of possible bin pairs, $r_H$ is the transition probability of the transition $H$, and $s_H=1-r_H$ is the remaining part of this probability division.
  Then, the probability division of the transition $(H-1)$ is conducted, and $s_H P(\bar{\bm{n}})$ is divided into $s_{H-1} P(\bar{\bm{n}})$ and $r_{H-1} P(\bar{\bm{n}})$.
  After performing all the probability divisions for every possible pair of bins, the probabilities at the next time level are obtained as indicated by the blue dots.}
  \label{fig:probability_division_single}
\end{figure*}

The probability division is conducted sequentially for all possible pairs of bins (Fig.~\ref{fig:probability_division_single}).
Each transition is labeled by $h$, and the corresponding transition probability from the state $\bar{\bm{n}}$ is written as $r_h(\bar{\bm{n}})$.
The label $h$ ranges from $0$ to $H$, where $H$ is the total number of possible bin pairs.
The correspondence between each bin pair and its transition label $h$ can be arbitrary, as long as it is uniquely defined.
We treat the case with no collisions, in which the state remains unchanged, separately.
This no-transition case is computed as the complement of all other transition probabilities, and is assigned the label $0$.
In addition to the transition probability $r_h(\bar{\bm{n}})$, we introduce the complement probability $s_{h}(\bar{\bm{n}})$ to describe the intermediate state of the system during the sequential execution of probability divisions.
It represents the fraction of the probability retained in the state $\bar{\bm{n}}$ after the probability division of the transition with label $h$, and satisfies the relationship 
\begin{equation}
  s_{h}(\bar{\bm{n}}) = 1-\sum_{k=h}^{H}r_k(\bar{\bm{n}}),
\end{equation}
by definition. Here, the index $k$ runs from $h$ to $H$ because the probability division is executed in descending order, starting from the maximum label $H$.
We introduce the modified transition probability $r_h'(\bar{\bm{n}})$ to calculate the transition probability $r_h(\bar{\bm{n}})$ from the remaining probability $s_{h+1}(\bar{\bm{n}})$.
This modified transition probability is defined as 
\begin{equation}
  r'_{h}(\bar{\bm{n}}) = \frac{r_{h}(\bar{\bm{n}})}{s_{h+1}(\bar{\bm{n}})}.
\end{equation}
Using these notations, as summarized in Fig.~\ref{fig:probability_division_single}, the sequence of probability-division operations within a single time step can be written as 
\begin{equation}
  \begin{aligned}
    P(\bar{\bm{n}}) \rightarrow &r_{H}P(\bar{\bm{n}}) + s_{H}P(\bar{\bm{n}}) \\
    \rightarrow &r_{H}P(\bar{\bm{n}}) + r'_{H-1}s_{H}P(\bar{\bm{n}}) + (1-r'_{H-1})s_{H}P(\bar{\bm{n}}) \\
    &= r_{H}P(\bar{\bm{n}}) + r_{H-1}P(\bar{\bm{n}}) + s_{H-1}P(\bar{\bm{n}}) \\
    \rightarrow &\cdots \\
    \rightarrow &r_{H}P(\bar{\bm{n}}) + r_{H-1}P(\bar{\bm{n}}) + \cdots + s_{1}P(\bar{\bm{n}}) \\
    &= \sum_{k=0}^{H}r_{k}P(\bar{\bm{n}}) \qquad (\because s_{1}=r_0),
  \end{aligned}
\end{equation}
where each arrow denotes one step of the probability-division procedure, corresponding to a single transition label.

\subsection{Quantum circuits}
In our quantum algorithm, a single time step is constructed by a sequential application of the quantum operations, which perform the probability divisions for each bin pair.
These quantum operations consist of four main components.
The first is the calculation of the probabilities $r'_h$ and $s_h$ using the fixed-point quantum arithmetic.
Its details are provided in Appendix~\ref{appendix:fixed_point}.
This component is constructed by the quantum gate $U_P(h)$.
The second is the application of the transition probability $r_h'$ to change the quantum amplitude, using controlled rotation gates.
This component is constructed by the quantum gate $U_{\sin}$.
The third is the uncomputation of the auxiliary qubits, which is constructed by $U_Q(h)$.
Here, ``uncomputation'' refers to applying the inverse of previously performed operations to return auxiliary qubits used in intermediate calculations to their initial states.
This process is required to reuse auxiliary qubits for repeated calculations because qubits cannot be reset directly unlike in classical computing.
The last is the update of the history register to keep track of the application order of probability divisions.
This component is constructed by the controlled adder gate $U_{\text{add}}$.
In the following paragraphs, we will explain the brief overview of the $U_P$, $U_{\sin}$, $U_Q$, and $U_{\text{add}}$ gates.
Refer to Appendix~\ref{appendix:quantum_algorithm_for_components} for the detailed explanations.

The $U_P(h)$ gate calculates the piecewise arcsine of the square root of the modified transition probability $r_h'$, i.e., $\arcsin \sqrt{r_h'}$.
This value gives the rotation angle of the rotation gate $R_y(\theta)$, which is applied inside the $U_{\sin}$ gate and expressed as
\begin{equation}
  R_y(\theta) = \begin{pmatrix}
    \cos{\frac{\theta}{2}} & -\sin{\frac{\theta}{2}} \\ \sin{\frac{\theta}{2}} & \cos{\frac{\theta}{2}}
  \end{pmatrix}\mbox{.}
\end{equation}
This gate acts as $R_y(2\arcsin{\sqrt{r'_{h}}})\ket{0} = \sqrt{1-r'_{h}}\ket{0}+\sqrt{r'_{h}}\ket{1}$.
The value of $r'_{h}$ can be computed if $i(h), j(h), n_{i(h)}, n_{j(h)}$ and  $s_{h+1}$ are given.
Here, $i(h)$ and $j(h)$ are the indices of the bins that collide in the transition with label $h$.
Therefore, the $U_P(h)$ gate acts as 
\begin{multline}
  \label{eq:U_P}
  U_P(h) : \ket{n_{i(h)}}_{\mathcal{N}_{i(h)}}\ket{n_{j(h)}}_{\mathcal{N}_{j(h)}}\ket{s_{h+1}}_{\mathcal{A}}\ket{0}_{\mathcal{B}}\ket{0}_{\mathcal{C}} \\ \mapsto \ket{n_{i(h)}}_{\mathcal{N}_{i(h)}}\ket{n_{j(h)}}_{\mathcal{N}_{j(h)}}\twocolbreak\ket{s_{h+1}}_{\mathcal{A}}\ket{\text{pp}_{\arcsin}{\sqrt{r'_h}}}_{\mathcal{B}}\ket{\lambda}_{\mathcal{C}},
\end{multline}
where $\ket{\lambda}$ is an auxiliary state used during the computation of $\text{pp}_{\arcsin}{\sqrt{r'_{h}}}$ and is a composite state comprising both integer and real numbers.
The symbol $\text{pp}_{\arcsin}$ means the piecewise polynomial to calculate arcsine~\cite{Haner2018-jy}.
$\mathcal{A}, \mathcal{B}$, and $\mathcal{C}$ are the quantum registers: $\mathcal{A}$ stores $s_{h+1}$, $\mathcal{B}$ stores $\text{pp}_{\arcsin}{\sqrt{r'_{h}}}$, and $\mathcal{C}$ stores auxiliary states. 
While the number of droplets in the $i$-th bin, $n_i$, is an integer number stored in the binary form in the qubit $\mathcal{N}_i$, the values of $s_{h+1}$ and $\text{pp}_{\arcsin}{\sqrt{r'_{h}}}$ are real numbers represented in the fixed-point form in the qubits $\mathcal{A}$ and $\mathcal{B}$, respectively.
These real numbers lie within the ranges $[0, 1]$ and $[0, \pi/2]$, respectively, and can be expressed with a precision of $n_{\epsilon}$ bits as
\begin{equation}
  \alpha = \alpha_{n_{\epsilon}-1}2^0 + \alpha_{n_{\epsilon}-2}2^{-1} + \cdots + \alpha_{0}2^{-n_{\epsilon}+1},
\end{equation}
where $\alpha$ represents either $s_{h+1}$ or $\text{pp}_{\arcsin}{\sqrt{r'_{h}}}$, $n_{\epsilon}$ denotes the number of qubits of this fixed-point representation, and $\alpha_{k} \in \{0,1\}\ (k=0, 1, \ldots, n_{\epsilon}-1)$.
The subscript $\epsilon$ of $n_{\epsilon}$ indicates that the number of qubits $n_{\epsilon}$ is determined by the target accuracy: the round-off error $2^{-n_{\epsilon}+1}$ of this representation contributes to the probability-calculation error $\epsilon_\text{calculation}$ defined in Sec.~\ref{Quantum resource estimates}.
This value is encoded in the quantum register as
\begin{equation}
  \ket{\alpha} = \ket{\alpha_{n_{\epsilon}-1}}\ket{\alpha_{n_{\epsilon}-2}}\ldots\ket{\alpha_{0}}.
\end{equation}
Here, we denote $\ket{1}\ket{0}\ldots\ket{0}$ by $\ket{1.0}$ for convenience.

The $U_{\sin}$ gate applies the probability division to the quantum amplitude.
The controlled rotation gates implement this operation.
The $U_{\sin}$ gate acts as 
\iftwocolumnversion
\begin{multline}
  \label{eq:U_sin}
  U_{\sin} : \ket{0}_{\mathcal{H}_m}\ket{\text{pp}_{\arcsin}{\sqrt{r'_{h}}}}_{\mathcal{B}}
  \\
  \mapsto \left[\sqrt{1-r'_{h}}\ket{0}_{\mathcal{H}_m} + \sqrt{r'_{h}}\ket{1}_{\mathcal{H}_m}\right]\ket{\text{pp}_{\arcsin}{\sqrt{r'_{h}}}}_{\mathcal{B}},
\end{multline}
\else
\begin{equation}
  \label{eq:U_sin}
  U_{\sin} : \ket{0}_{\mathcal{H}_m}\ket{\text{pp}_{\arcsin}{\sqrt{r'_{h}}}}_{\mathcal{B}}
  \mapsto \left[\sqrt{1-r'_{h}}\ket{0}_{\mathcal{H}_m} + \sqrt{r'_{h}}\ket{1}_{\mathcal{H}_m}\right]\ket{\text{pp}_{\arcsin}{\sqrt{r'_{h}}}}_{\mathcal{B}},
\end{equation}
\fi
where $\mathcal{H}_m$ is the quantum register storing the transition label at the $m$-th time step. 
Notably, the $U_{\sin}$ gate is applied only when the quantum register $\mathcal{H}_m$ is in the $\ket{0}_{\mathcal{H}_m}$ state. This means that the probability-division operation is applied only to a fraction of the probability that has not been affected by any previous transition labels within the $m$-th time step.

After completing the probability division for the transition with label $h$, we uncompute the auxiliary states of the qubits $\mathcal{B}$ and $\mathcal{C}$, and update the value of $s_{h+1}$ to $s_{h}$, which is used in the probability division in the next transition $(h-1)$. The $U_Q(h)$ gate achieves this operation as 
\begin{multline}
  \label{eq:U_Q}
  U_Q(h) : \ket{n_{i(h)}}_{\mathcal{N}_{i(h)}}\ket{n_{j(h)}}_{\mathcal{N}_{j(h)}}\twocolbreak\ket{s_{h+1}}_{\mathcal{A}}\ket{\text{pp}_{\arcsin}{\sqrt{r'_{h}}}}_{\mathcal{B}}\ket{\lambda}_{\mathcal{C}} \\ \mapsto \ket{n_{i(h)}}_{\mathcal{N}_{i(h)}}\ket{n_{j(h)}}_{\mathcal{N}_{j(h)}}\ket{s_{h}}_{\mathcal{A}}\ket{0}_{\mathcal{B}}\ket{0}_{\mathcal{C}}.
\end{multline}

Following the uncomputation, the $U_{\text{add}}$ gate is applied on the quantum register $\mathcal{H}_m$.
Specifically, this operation maps $\ket{h}_{\mathcal{H}_m}$ to $\ket{h+1}_{\mathcal{H}_m}$ only when $h\geq 1$, while $\ket{0}_{\mathcal{H}_m}$ remains unchanged.
This operation is performed to keep track of whether the probability-division operation has already been applied to a given state and to record the transition label that has been used.

As noted earlier, the probability divisions are applied in descending order from $h=H$ down to $h=1$.
When a state is first updated at index $h$, its history register is set to 1.
In each subsequent step, the register is incremented by 1.
Therefore, once the probability-division procedure has been finished, the final value of the history register becomes equal to the transition label at which the state was updated.

\subsection[Quantum circuit for a single time step]{Quantum circuit for a single time step}
The quantum circuit to perform the probability division for a single time step is explained.
The following example is the case of applying the transition with label $h$ at the $m$-th time step.
Just before this probability-division operation, the quantum state is 
\begin{multline}
  \label{eq:before_transition_h}
  \sum_{h_1,\ldots,h_{m-1}}\sum_{\bar{\bm{n}}}\ket{h_1}_{\mathcal{H}_1}\ldots\ket{h_{m-1}}_{\mathcal{H}_{m-1}}
  \\
  \left[\sum_{k=h+1}^{H}\sqrt{r_{k}P(\bm{h},\bar{\bm{n}};(m-1)\Delta t)}\ket{k-h+1}_{\mathcal{H}_m}
  \iftwocolumnversion\right. \\ \left.\fi
  + \sqrt{s_{h+1}P(\bm{h},\bar{\bm{n}};(m-1)\Delta t)}\ket{0}_{\mathcal{H}_m}\right]
  \\
  \ket{0}_{\mathcal{H}_{m+1}}\ldots\ket{0}_{\mathcal{H}_{M}}\ket{\bar{\bm{n}}}_{\mathcal{N}}\ket{s_{h+1}}_{\mathcal{A}}.
\end{multline}
By sequentially applying the $U_P(h), U_{\sin}$, and $U_Q(h)$ gates defined by Eqs.~\eqref{eq:U_P}, \eqref{eq:U_sin}, and \eqref{eq:U_Q}, respectively, the probability division for the transition with label $h$ transforms the quantum state Eq.~\eqref{eq:before_transition_h} into the following:
\begin{equation}
  \iftwocolumnversion
  \begin{aligned}
    &\sum_{h_1,\ldots,h_{m-1}}\sum_{\bar{\bm{n}}}\ket{h_1}_{\mathcal{H}_1}\ldots\ket{h_{m-1}}_{\mathcal{H}_{m-1}}
    \\ &
    \left[\sum_{k=h+1}^{H}\sqrt{r_{k}P(\bm{h},\bar{\bm{n}};(m-1)\Delta t)}\ket{k-h+1}_{\mathcal{H}_m} \right. \\
    &+ \sqrt{r_h's_{h+1}P(\bm{h},\bar{\bm{n}};(m-1)\Delta t)}\ket{1}_{\mathcal{H}_m}
    \\ &
    \left. + \sqrt{(1-r_h')s_{h+1}P(\bm{h},\bar{\bm{n}};(m-1)\Delta t)}\ket{0}_{\mathcal{H}_m}\right] \\
    &\qquad \qquad \ket{0}_{\mathcal{H}_{m+1}}\ldots\ket{0}_{\mathcal{H}_{M}}\ket{\bar{\bm{n}}}_{\mathcal{N}}\ket{s_{h}}_{\mathcal{A}} \\
    =&\sum_{h_1,\ldots,h_{m-1}}\sum_{\bar{\bm{n}}}\ket{h_1}_{\mathcal{H}_1}\ldots\ket{h_{m-1}}_{\mathcal{H}_{m-1}}
    \\ &
    \left[\sum_{k=h}^{H}\sqrt{r_{k}P(\bm{h},\bar{\bm{n}};(m-1)\Delta t)}\ket{k-h+1}_{\mathcal{H}_m} \right. \\
    &\left. + \sqrt{s_{h}P(\bm{h},\bar{\bm{n}};(m-1)\Delta t)}\ket{0}_{\mathcal{H}_m}\right]
    \\ & \qquad \qquad
    \ket{0}_{\mathcal{H}_{m+1}}\ldots\ket{0}_{\mathcal{H}_{M}}\ket{\bar{\bm{n}}}_{\mathcal{N}}\ket{s_{h}}_{\mathcal{A}}.
  \end{aligned}
  \else
  \begin{aligned}
    &\sum_{h_1,\ldots,h_{m-1}}\sum_{\bar{\bm{n}}}\ket{h_1}_{\mathcal{H}_1}\ldots\ket{h_{m-1}}_{\mathcal{H}_{m-1}}
    \left[\sum_{k=h+1}^{H}\sqrt{r_{k}P(\bm{h},\bar{\bm{n}};(m-1)\Delta t)}\ket{k-h+1}_{\mathcal{H}_m} \right. \\
    &\qquad + \sqrt{r_h's_{h+1}P(\bm{h},\bar{\bm{n}};(m-1)\Delta t)}\ket{1}_{\mathcal{H}_m}
    \left. + \sqrt{(1-r_h')s_{h+1}P(\bm{h},\bar{\bm{n}};(m-1)\Delta t)}\ket{0}_{\mathcal{H}_m}\right] \\
    &\qquad \qquad \ket{0}_{\mathcal{H}_{m+1}}\ldots\ket{0}_{\mathcal{H}_{M}}\ket{\bar{\bm{n}}}_{\mathcal{N}}\ket{s_{h}}_{\mathcal{A}} \\
    =&\sum_{h_1,\ldots,h_{m-1}}\sum_{\bar{\bm{n}}}\ket{h_1}_{\mathcal{H}_1}\ldots\ket{h_{m-1}}_{\mathcal{H}_{m-1}}
    \left[\sum_{k=h}^{H}\sqrt{r_{k}P(\bm{h},\bar{\bm{n}};(m-1)\Delta t)}\ket{k-h+1}_{\mathcal{H}_m} \right. \\
    &\qquad \left. + \sqrt{s_{h}P(\bm{h},\bar{\bm{n}};(m-1)\Delta t)}\ket{0}_{\mathcal{H}_m}\right]
    \ket{0}_{\mathcal{H}_{m+1}}\ldots\ket{0}_{\mathcal{H}_{M}}\ket{\bar{\bm{n}}}_{\mathcal{N}}\ket{s_{h}}_{\mathcal{A}}.
  \end{aligned}
  \fi
\end{equation}
Subsequently, the $U_{\text{add}}$ gate is applied to increment the history register by one.
As a result, the resulting quantum state becomes 
\begin{multline}
  \sum_{h_1,\ldots,h_{m-1}}\sum_{\bar{\bm{n}}}\ket{h_1}_{\mathcal{H}_1}\ldots\ket{h_{m-1}}_{\mathcal{H}_{m-1}}
  \\
  \left[\sum_{k=h}^{H}\sqrt{r_{k}P(\bm{h},\bar{\bm{n}};(m-1)\Delta t)}\ket{k-h+2}_{\mathcal{H}_m}
  \iftwocolumnversion\right. \\ \left.\fi
  + \sqrt{s_{h}P(\bm{h},\bar{\bm{n}};(m-1)\Delta t)}\ket{0}_{\mathcal{H}_m}\right]
  \\
  \ket{0}_{\mathcal{H}_{m+1}}\ldots\ket{0}_{\mathcal{H}_{M}}\ket{\bar{\bm{n}}}_{\mathcal{N}}\ket{s_{h}}_{\mathcal{A}}.
\end{multline}

The series of these four components explained above is sequentially applied for all of the possible bin pairs.
After completing all probability divisions in the $m$-th time step, the quantum register $\mathcal{A}$ holds the value $\ket{s_1}_{\mathcal{A}}$.
To reuse $\mathcal{A}$ in the next time step, $\ket{s_1}_{\mathcal{A}}$ is uncomputed to the initial state $\ket{1.0}_{\mathcal{A}}$.
This is the inverse of the combined operations of $U_Q(h)$ and $U_P(h)$, and is achieved by iteratively applying the $U_R(h)$ gate.
Additionally, the states that have non-zero transition labels require an update of the mass distribution to the post-collision state.
For example, if a transition label is for a collision between droplets in the first bin and the current state is $\ket{N}_{\mathcal{N}_1}\ket{0}_{\mathcal{N}_2}\ldots\ket{0}_{\mathcal{N}_N}$, in which the first bin holds all $N$ droplets as in Eq.~\eqref{eq:initial_state}, the state must be updated to $\ket{N-2}_{\mathcal{N}_1}\ket{1}_{\mathcal{N}_2}\ldots\ket{0}_{\mathcal{N}_N}$.
This update is performed using the $U_{\text{shift}}$ gate, which consists of constant additions and subtractions.
Details of $U_R(h)$ and $U_{\text{shift}}$ gates are also explained in Appendix~\ref{appendix:quantum_algorithm_for_components}.

The complete set of operations for a single time step is summarized in the quantum circuit in Fig.~\ref{fig:qcirc_one_step}.
The operator for calculating a single time step is denoted as $U_{\Delta t}$ on the left-hand-side.
This operates such that $U_{\Delta t}\ket{\psi((m-1)\Delta t)}_{\mathcal{HN}}\ket{1.0}_{\mathcal{A}}\ket{0}_{\mathcal{B}}\ket{0}_{\mathcal{C}}=\ket{\psi(m\Delta t)}_{\mathcal{HN}}\ket{1.0}_{\mathcal{A}}\ket{0}_{\mathcal{B}}\ket{0}_{\mathcal{C}}$. 

\begin{figure*}
  \resizebox{\textwidth}{!}{
    \begin{quantikz}[transparent, row sep={20pt,between origins}, column sep=4pt]
      \lstick{$\ket{0}_{\mathcal{H}_m}$} & \gate[5]{U_{\Delta t}} & \qw & \rstick[2]{$\sum_{h,\bar{\bm{n}}'}\sqrt{r_h}\ket{h,\bar{\bm{n}}'}_{\mathcal{H}_m,\mathcal{N}}$} \\
      \lstick{$\ket{\bar{\bm{n}}}_{\mathcal{N}}$} & & \qw & \\
      \lstick{$\ket{1.0}_{\mathcal{A}}$} & & \qw & \rstick{$\ket{1.0}_{\mathcal{A}}$} \\
      \lstick{$\ket{0}_{\mathcal{B}}$} & & \qw & \rstick{$\ket{0}_{\mathcal{B}}$} \\
      \lstick{$\ket{0}_{\mathcal{C}}$} & & \qw & \rstick{$\ket{0}_{\mathcal{C}}$} \\
    \end{quantikz}
  = 
    \begin{quantikz}[transparent, row sep={20pt,between origins}, column sep=8pt]
      & \qw \gategroup[5,steps=7, style={dashed,inner sep=0.3pt,blue}, label style={label position=below, yshift=-0.5cm}]{} & \gate[4,label style={yshift=-25pt}]{U_{\sin}} & \gate{U_{\text{add}}} & \qw \ \ldots \ & \qw & \gate[4,label style={yshift=-25pt}]{U_{\sin}} & \qw & \qw & \qw & \qw & \gate[2]{U_\text{shift}(H)} \gategroup[2,steps=3, style={dashed,inner sep=-0.5pt,blue}, label style={label position=below, yshift=-0.5cm}]{} & \qw \ \ldots \ & \gate[2]{U_\text{shift}(1)} & \qw \\
      & \gate[4]{U_P(H)} & \linethrough & \gate[4]{U_Q(H)} & \qw \ \ldots \ & \gate[4]{U_P(1)} & \linethrough & \gate[4]{U_Q(1)} & \gate[4]{U_R(H)} \gategroup[4,steps=3, style={dashed,inner sep=-0.5pt,blue}, label style={label position=below, yshift=-20pt,blue}]{Repeat $H$ times from $H$ to $1$} & \qw \ \ldots \ & \gate[4]{U_R(1)} &  & \qw \ \ldots \ & & \qw \\
      & \qw & \linethrough & & \qw \ \ldots \ & \qw & \linethrough & & & \qw \ \ldots \ & & \qw & \qw & \qw & \qw \\
      &  & \qw & & \qw \ \ldots \ &  & \qw & & & \qw \ \ldots \ & & \qw & \qw & \qw & \qw \\
      & \qw & \qw & & \qw \ \ldots \ & \qw & \qw & & & \qw \ \ldots \ & & \qw & \qw & \qw & \qw 
    \end{quantikz}
  }
  \caption{The quantum circuit for a single time step.
  The left-hand-side of the equation is the input and output states of this time step, while the right-hand-side illustrates the detailed operations within this time step.
  These segments enclosed by blue dashed lines on the right-hand-side include sequential applications of probability-division operations for all possible bin pairs (omitted for simplicity, represented by dots in the figure).
  The left segment calculates the modified transition probability and applies the probability-division operations to the quantum amplitude.
  The central segment uncomputes the auxiliary qubits $\mathcal{A}$ to $\ket{1.0}_\mathcal{A}$.
  The right segment updates the mass distribution $\ket{\bar{\bm{n}}}_{\mathcal{N}}$ to the next time level based on the transition label $\ket{h}_{\mathcal{H}_m}$.
  The mass distribution obtained after the transition with label $h$ is denoted by $\bar{\bm{n}}'$.}
  \label{fig:qcirc_one_step} 
\end{figure*}

\begin{figure}[htbp]
  \centering
  \resizebox{\columnwidth}{!}{
  \begin{quantikz}[transparent, row sep={15pt,between origins}, column sep=5pt]
    \lstick{$\ket{0}_{\mathcal{H}_1}$} & \gate[7,nwires={2,4}]{U_t} & \qw & \rstick[6]{$\ket{\psi(t)}_{\mathcal{HN}}$} \\
    & & & \\
    \lstick{$\ket{0}_{\mathcal{H}_m}$} & & \qw & \\
    & & & \\
    \lstick{$\ket{0}_{\mathcal{H}_M}$} & & \qw & \\
    \lstick{$\ket{\bm{0}}_{\mathcal{N}}$} & & \qw & \\
    \lstick{$\ket{\bm{0}}_{\mathcal{A}\mathcal{B}\mathcal{C}}$} & & \qw & \rstick{$\ket{1.0}_{\mathcal{A}}\ket{\bm{0}}_{\mathcal{B}\mathcal{C}}$} 
  \end{quantikz}
  =
  \begin{quantikz}[transparent, row sep={15pt,between origins}, column sep=8pt]
    & \qw & \gate[7,nwires={2,4},label style={yshift=45pt}]{U_{\Delta t}} \gategroup[7,steps=5, style={inner sep=1pt}, label style={label position=below, yshift=-0.5cm}]{$M$ times} & \qw & \qw & \qw & \qw & \qw & \\
    & & & \ddots & & & & \\
    & \qw & \linethrough & \qw & \gate[5,nwires={2},label style={yshift=25pt}]{U_{\Delta t}} & \qw & \qw & \qw & \\
    & & & & & \ddots & & \\
    & \qw & \linethrough & \qw & \linethrough & \qw & \gate[3]{U_{\Delta t}} & \qw & \\
    & \gate[2]{\text{INIT}} &  &\qw\ \ldots \ & &\qw\ \ldots \ & & \qw & \\
    & &  &\qw\ \ldots \ & &\qw\ \ldots \ & & \qw &
  \end{quantikz}
  }
  \caption{The quantum circuit for calculating the time evolution of the probability distribution over all time steps.
  The circuit begins with the initialization of the droplet mass distribution and the auxiliary qubits $\mathcal{A}$ in the state $\ket{1.0}_\mathcal{A}$.
  Subsequently, the quantum gates $U_{\Delta t}$ are sequentially applied $M$ times. The history register $\mathcal{H}_m$ is used for the $m$-th time step.}
  \label{fig:qcirc_all_step}
\end{figure}

\subsection[Quantum circuit for all time steps]{Quantum circuit for all time steps}
The quantum circuit for the entire time evolution is shown in Fig.~\ref{fig:qcirc_all_step}.
The process begins by initializing the quantum registers to the states of the initial conditions. This initialization is performed by simply applying NOT gates.
After the initialization, the probability division for each time step is carried out sequentially.
As the time proceeds, the target qubits of the controlled rotation gate change from $\mathcal{H}_1$ to $\mathcal{H}_M$.
The operator for calculating the final state is denoted as $U_t$ on the left-hand-side.
This operates such that $U_t\ket{0}_{\mathcal{HN}}\ket{0}_{\mathcal{A}}\ket{0}_{\mathcal{B}}\ket{0}_{\mathcal{C}}=\ket{\psi(t)}_{\mathcal{HN}}\ket{1.0}_{\mathcal{A}}\ket{0}_{\mathcal{B}}\ket{0}_{\mathcal{C}}$.

\subsection{Quantum circuit for postprocessing}
After the time evolution is finished, we derive characteristic values from the quantum state $\ket{\psi(t)}_{\mathcal{HN}}$ at the final step.
Since the quantum state is expressed as in Eqs.~\eqref{eq:quantum_state}, \eqref{eq:transition_history}, and \eqref{eq:mass_distribution}, the characteristic values can be obtained efficiently.
For example, the probability of observing $n_i=n$ in Eq.~\eqref{eq:probability} is encoded directly in the quantum amplitude of $\ket{n}_{\mathcal{N}_i}$.
This can be verified as
\begin{align}
  \iftwocolumnversion
  &P(n_i=n;t) \nonumber\\
  &= \sum_{\bm{h}}\sum_{\{n_j\}_{j \neq i}}P(\bm{h},n_1,\ldots,n_i=n,\ldots,n_N;t) \nonumber\\
  &= \left\lVert\sum_{\bm{h}}\sum_{\{n_j\}_{j \neq i}}\sqrt{P(\bm{h},n_1,\ldots,n_i=n,\ldots,n_N;t)}
  \right. \nonumber\\ &\left. \qquad
  \ket{\bm{h}}_{\mathcal{H}}\ket{n_1}_{\mathcal{N}_1}\ldots\ket{n}_{\mathcal{N}_i}\ldots\ket{n_N}_{\mathcal{N}_N}\right\rVert^2 \nonumber\\
  &= \left\lVert I_{\setminus \mathcal{N}_i}\otimes\bra{n}_{\mathcal{N}_i}\ket{\psi(t)}_{\mathcal{HN}}\right\rVert^2,
  \else
  P(n_i=n;t)
  &= \sum_{\bm{h}}\sum_{\{n_j\}_{j \neq i}}P(\bm{h},n_1,\ldots,n_i=n,\ldots,n_N;t) \nonumber\\
  &= \left\lVert\sum_{\bm{h}}\sum_{\{n_j\}_{j \neq i}}\sqrt{P(\bm{h},n_1,\ldots,n_i=n,\ldots,n_N;t)}
  \right. \nonumber\\ &\left. \qquad
  \ket{\bm{h}}_{\mathcal{H}}\ket{n_1}_{\mathcal{N}_1}\ldots\ket{n}_{\mathcal{N}_i}\ldots\ket{n_N}_{\mathcal{N}_N}\right\rVert^2 \nonumber\\
  &= \left\lVert I_{\setminus \mathcal{N}_i}\otimes\bra{n}_{\mathcal{N}_i}\ket{\psi(t)}_{\mathcal{HN}}\right\rVert^2,
  \fi
\end{align}
where $I_{\setminus \mathcal{N}_i}$ is the identity operator acting on all qubits except $\mathcal{N}_i$.
We can use the quantum amplitude estimation~\cite{Brassard-ut,Grinko2021-nk} for an efficient estimation of $P(n_i=n;t)$. 

The expected values of the number of droplets in the $i$-th bin defined in Eq.~\eqref{eq:expected_value} can also be estimated by the quantum amplitude estimation.
The expected values can be calculated by applying the controlled rotation gate $U_c(i)$ sequentially.
The $U_c(i)$ is defined to encode the number of droplets in the $i$-th bin into the quantum amplitude as follows:
\iftwocolumnversion
\begin{multline}
  \label{eq:U_c}
  U_c(i) : \ket{0}_{\mathcal{D}}\ket{n_i}_{\mathcal{N}_i} \\
  \mapsto \left[\sqrt{\frac{n_i}{d}}\ket{0}_{\mathcal{D}}+\sqrt{1-\frac{n_i}{d}}\ket{1}_{\mathcal{D}}\right]\ket{n_i}_{\mathcal{N}_i},
\end{multline}
\else
\begin{equation}
  \label{eq:U_c}
  U_c(i) : \ket{0}_{\mathcal{D}}\ket{n_i}_{\mathcal{N}_i} \mapsto \left[\sqrt{\frac{n_i}{d}}\ket{0}_{\mathcal{D}}+\sqrt{1-\frac{n_i}{d}}\ket{1}_{\mathcal{D}}\right]\ket{n_i}_{\mathcal{N}_i},
\end{equation}
\fi
where $d$ is a value larger than the maximum number of droplets in the $i$-th bin, and $\mathcal{D}$ is an auxiliary qubit. In this work, we use $d=2^{q_i}$, where $q_i$ denotes the number of qubits required to store the quantum state $\ket{n_i}_{\mathcal{N}_i}$.
The detailed algorithm for the construction of $U_c$ is explained in Appendix~\ref{appendix:quantum_algorithm_for_components}. 

Figure~\ref{fig:qcirc_expected_value} shows the quantum circuit that produces the final state expressed as
\iftwocolumnversion
\begin{multline}
  \ket{\psi_{\text{final}}}_{\mathcal{DHN}}
  = \sum_{\bm{h}}\sum_{\bar{\bm{n}}}\left[\sqrt{\frac{n_i}{d}}\ket{0}_{\mathcal{D}}+\sqrt{1-\frac{n_i}{d}}\ket{1}_{\mathcal{D}}\right]
  \\
  \sqrt{P(\bm{h},\bar{\bm{n}};t)}\ket{\bm{h}}_{\mathcal{H}}\ket{\bar{\bm{n}}}_{\mathcal{N}}.
\end{multline}
\else
\begin{equation}
  \ket{\psi_{\text{final}}}_{\mathcal{DHN}}
  = \sum_{\bm{h}}\sum_{\bar{\bm{n}}}\left[\sqrt{\frac{n_i}{d}}\ket{0}_{\mathcal{D}}+\sqrt{1-\frac{n_i}{d}}\ket{1}_{\mathcal{D}}\right]
  \sqrt{P(\bm{h},\bar{\bm{n}};t)}\ket{\bm{h}}_{\mathcal{H}}\ket{\bar{\bm{n}}}_{\mathcal{N}}.
\end{equation}
\fi
Therefore, the expected value of the number of droplets in the $i$-th bin can be obtained by measuring the probability of finding the quantum state $\ket{0}_{\mathcal{D}}$ as
\begin{align}
  \nonumber
  \braket{n_i} &= d\sum_{n}\frac{n}{d}P(n_i=n;t) \\
  \nonumber
  &= d\left\lVert\sum_{n}\sqrt{\frac{n}{d}}\sum_{\bm{h}}\sum_{\{n_j\}_{j \neq i}}\sqrt{P(\bm{h},\bar{\bm{n}};t)}\ket{\bm{h}}_{\mathcal{H}}\ket{\bar{\bm{n}}}_{\mathcal{N}}\right\rVert^2 \\
  &= d\left\lVert I_{\mathcal{HN}}\otimes\bra{0}_{\mathcal{D}}\ket{\psi_{\text{final}}}_{\mathcal{DHN}}\right\rVert^2,
  \label{eq:expected_value_amplitude}
\end{align}
and the quantum amplitude estimation allows an efficient estimation of this value.

\begin{figure}[htbp]
  \centering
  \begin{quantikz}[transparent, row sep={20pt,between origins}, column sep=20pt]
    \lstick{$\ket{0}_{\mathcal{D}}$} & \qw & \gate[5,nwires={4},label style={yshift=-25pt}]{U_c} & \qw  \\
    \lstick{$\ket{\bm{0}}_{\mathcal{H}}$} & \gate[7,nwires={3,5}]{U_t} & \linethrough & \qw \\
    \lstick{$\ket{0}_{\mathcal{N}_1}$} & & \linethrough & \qw\\
    & & & \\
    \lstick{$\ket{0}_{\mathcal{N}_i}$} &  & & \qw \\
    & & & \\
    \lstick{$\ket{0}_{\mathcal{N}_N}$} & & \qw & \qw \\
    \lstick{$\ket{\bm{0}}_{\mathcal{A}\mathcal{B}\mathcal{C}}$} & & \qw & \qw
  \end{quantikz}
  \caption{The quantum circuit for calculating the expected value of the number of droplets in the $i$-th bin. The expected value is encoded into the quantum amplitude of the auxiliary qubit $\mathcal{D}$.}
  \label{fig:qcirc_expected_value}
\end{figure}

\section{\label{Quantum resource estimates}Quantum resource estimates}
To assess the feasibility of our algorithm on fault-tolerant quantum computers, we estimate its quantum resources in two respects: (i) the logical-qubit count, including the main registers encoding $\ket{\psi(t)}$ and the auxiliary registers, and (ii) the number of T gates (T-count). We focus on the T gate because it requires significantly more computational time than the Clifford gates~\cite{Goings2022-wr}.

First, we count the number of logical qubits required to represent the quantum state $\ket{\psi(t)}$.
The mathematical expression of $\ket{\psi(t)}$ is given by Eqs.~\eqref{eq:quantum_state}, \eqref{eq:transition_history}, and \eqref{eq:mass_distribution}.
In our problem setup, the total mass existing in the system is always equal to the largest possible droplet mass $x_N$, because we take the equality in the condition $x_\text{total} \leq x_N$ introduced in Sec.~\ref{Formulation}.
Thus, the maximum number of droplets in the $i$-th bin is given by $n_i^{\max} = \lfloor N/i \rfloor$, where $\lfloor * \rfloor$ is the greatest integer less than or equal to $*$.
The number of logical qubits required for $\ket{n_i}_{\mathcal{N}_i}$ is calculated as $q_i = \lceil \log_2(\lfloor N/i \rfloor+1) \rceil$, where $\lceil * \rceil$ is the smallest integer greater than or equal to $*$.
Consequently, the total number of logical qubits required for the quantum state $\ket{\bar{\bm{n}}}_{\mathcal{N}}$ is given by
\begin{equation}
  q_{\text{count}}(\mathcal{N}) = \sum_{i=1}^N \lceil \log_2(\lfloor N/i \rfloor+1) \rceil,
\end{equation}
where $q_{\text{count}}(*)$ denotes the number of logical qubits required for the quantum register $*$.
The number of logical qubits required for the quantum state $\ket{h_m}_{\mathcal{H}_m}$ can be calculated from the total number of possible bin pairs $H$ introduced in Sec.~\ref{Quantum algorithm}. From Eq.~\eqref{eq:master_discretized} and the conservation of total mass, 
\begin{equation}
  \label{eq:H_count}
  \iftwocolumnversion
  H = \begin{cases}
    \begin{aligned}
      &\sum_{i=1}^{N/2-1}\sum_{j=i+1}^{N-i} 1 + \sum_{i=1}^{N/2} 1 \\
      &= \frac{N^2}{4}
    \end{aligned} & (N: \text{even}), \\
    \begin{aligned}
      &\sum_{i=1}^{(N-1)/2}\sum_{j=i+1}^{N-i} 1 + \sum_{i=1}^{(N-1)/2} 1 \\
      &= \frac{N^2-1}{4}
    \end{aligned} & (N: \text{odd}).
  \end{cases}
  \else
  H = \begin{cases}
    \sum_{i=1}^{N/2-1}\sum_{j=i+1}^{N-i} 1 + \sum_{i=1}^{N/2} 1 = \frac{N^2}{4} & (N: \text{even}), \\
    \sum_{i=1}^{(N-1)/2}\sum_{j=i+1}^{N-i} 1 + \sum_{i=1}^{(N-1)/2} 1 = \frac{N^2-1}{4} & (N: \text{odd}).
  \end{cases}
  \fi
\end{equation}
The number of logical qubits required for $\ket{h_m}_{\mathcal{H}_m}$ is $q_h = \lceil \log_2(H+1) \rceil$, which is invariant across all time steps.
Therefore, the total number of logical qubits required for the quantum state $\ket{\bm{h}}_{\mathcal{H}}$ can be calculated as 
\begin{equation}
  q_\text{count}(\mathcal{H}) = M \lceil \log_2(H+1) \rceil.
\end{equation}

The logical-qubit count also includes the auxiliary qubits used in the quantum circuit. The numbers of qubits required for the auxiliary registers $\mathcal{A}, \mathcal{B}, \mathcal{C}$, and $\mathcal{D}$ are:
\begin{align}
  q_{\text{count}}(\mathcal{A}) &= n_{\epsilon}, \\
  q_{\text{count}}(\mathcal{B}) &= n_{\epsilon}, \\
  q_{\text{count}}(\mathcal{C}) &= 4q_1 + 5n_{\epsilon} + 1, \text{and} \\
  q_{\text{count}}(\mathcal{D}) &= 1,
\end{align}
where $n_{\epsilon}$ is the number of qubits of the fixed-point representation introduced in Sec.~\ref{Quantum algorithm}, and $q_1$ is the number of qubits for the first bin.
The $4q_1$ term in $q_{\text{count}}(\mathcal{C})$ comes from the registers $\mathcal{C}_1$, $\mathcal{C}_2$, and $\mathcal{C}_3$ of Appendix~\ref{appendix:quantum_algorithm_for_components}, which hold $n_{i(h)}$, $n_{j(h)}$, and their product with $q_1$, $q_1$, and $2q_1$ qubits, respectively.
Additionally, auxiliary qubits for quantum arithmetic operations must also be taken into account.
The details of the counts of these auxiliary qubits are explained in Appendix~\ref{appendix:quantum_algorithm_for_components}.

Second, we count the number of logical quantum gates. A single application of the composite operator $A$ shown in Fig.~\ref{fig:concept} consists of state preparation, time evolution, and expected-value calculation. We first estimate the T-count of each component and then multiply by the number of applications of $A$ determined by quantum amplitude estimation.

The initial condition is deterministic since all droplets reside in the first bin, as in Eq.~\eqref{eq:initial_state}. This is the canonical setup in collision-coalescence studies~\cite{Alfonso2015-vq}. Therefore, the initial state $\ket{\psi(0)}$ corresponds to a single computational-basis state, and can be prepared from $\ket{\bm{0}}$ by applying at most $q_1$ NOT gates to the register of the first bin. Since the NOT gate is a Clifford gate, state preparation needs zero T gates. Preparation of a general initial distribution would require a dedicated state-preparation circuit, but this is not the case in this problem.

For the time-evolution part, we count the number of logical quantum gates for a single operation of $U_P, U_{\sin}, U_Q, U_{\text{add}}, U_R$, and $U_{\text{shift}}$, respectively, and multiply them by the number of times each operation is called.
The detailed algorithms and the resources are provided in Appendix~\ref{appendix:quantum_algorithm_for_components}, and the summary of the T-count for each operation is given below:
\begin{align}
  \nonumber
  T_\text{count}(U_P) =& T_\text{count}(\text{MUL\_INT}_{q_1,q_1}) 
  \\ \nonumber &
  + T_\text{count}(\text{MUL\_CONST\_INT\_UI}_{2q_1,n_{\epsilon}}) \\ \nonumber&+ T_\text{count}(\text{COMP}_{n_{\epsilon}}) 
  \\ \nonumber &
  + 2T_\text{count}(\text{c-SUB}_{n_{\epsilon}}) \\ \nonumber&+ 2T_\text{count}(\text{SQRT}_{n_{\epsilon}}) + T_\text{count}(\text{DIV}_{n_{\epsilon}}) \\ &+ T_\text{count}(\text{ARCSIN}_{n_{\epsilon},\epsilon_{\arcsin}}), \\
  T_\text{count}(U_{\sin}) =& 12n_{\epsilon}+6.6\log_2(4/\epsilon_{\sin}) + 8q_h - 16, \\
  \nonumber
  T_\text{count}(U_Q) =& T_\text{count}(\text{MUL\_INT}_{q_1,q_1}) 
  \\ \nonumber &
  + T_\text{count}(\text{MUL\_CONST\_INT\_UI}_{2q_1,n_{\epsilon}}) \\ \nonumber&+ T_\text{count}(\text{COMP}_{n_{\epsilon}}) 
  \\ \nonumber &
  + 2T_\text{count}(\text{c-SUB}_{n_{\epsilon}}) \\ \nonumber&+ 2T_\text{count}(\text{SQRT}_{n_{\epsilon}}) + T_\text{count}(\text{DIV}_{n_{\epsilon}}) \\ &+ T_\text{count}(\text{ARCSIN}_{n_{\epsilon},\epsilon_{\arcsin}}) 
  \\ \nonumber &
  + T_\text{count}(\text{SUB}_{n_{\epsilon}}), \\
  T_\text{count}(U_{\text{add}}) =& T_\text{count}(\text{ADD\_CONST}_{q_h}) 
  \\ \nonumber &
  + T_\text{count}(\text{Toffoli}_{q_h}), \\
  \nonumber
  T_\text{count}(U_R) =& 2T_\text{count}(\text{MUL\_INT}_{q_1,q_1}) 
  \\ \nonumber &
  + 2T_\text{count}(\text{MUL\_CONST\_INT\_UI}_{2q_1,n_{\epsilon}}) \\ &+ T_\text{count}(\text{ADD}_{n_{\epsilon}}), \\
  \label{eq:T_count_U_shift}
  T_\text{count}(U_{\text{shift}}) =& \begin{cases}
    2T_\text{count}(\text{Toffoli}_{q_h}) \twocolbreak + T_\text{count}(\text{c-ADD}_{q_{i+j}})
    \\
    + T_\text{count}(\text{c-SUB}_{q_{i}}) \twocolbreak + T_\text{count}(\text{c-SUB}_{q_{j}}) \\
    \phantom{+ T_\text{count}(\text{c-SUB}_{q_{i}}) \iftwocolumnversion\qquad\else\qquad \qquad\fi}\text{(case 1)}, \\
    2T_\text{count}(\text{Toffoli}_{q_h}) \twocolbreak + T_\text{count}(\text{c-ADD}_{q_{2i}})
    \\
    + T_\text{count}(\text{c-SUB}_{q_{i}}) \iftwocolumnversion\qquad\else\qquad \qquad\fi \text{(case 2)},
  \end{cases}
\end{align}
where $T_\text{count}(*)$ denotes the T-count of the operation $*$.
Here, $\epsilon_{\sin}$ is the error tolerance of the controlled rotations inside $U_{\sin}$, and $\epsilon_{\arcsin}$ is the error tolerance of the piecewise-polynomial approximation of the arcsine inside $U_P$ and $U_Q$.
The value of $\epsilon_{\arcsin}$ determines the parameters $d_{\epsilon}$ and $S_{\epsilon}$ of the piecewise polynomial through Table~\ref{table:piecewise_arcsin}.
In Eq.~\eqref{eq:T_count_U_shift}, the T-count for $U_\text{shift}$ depends on the pair of bins in a collision.
In case 1, the transition is due to a collision between a pair of droplets of different masses, whereas in case 2, it is due to a collision between a pair of droplets of equal mass.
In the operation of $U_{\Delta t}$, the number of calls for each component except for the $U_{\text{add}}$ gate corresponds to the total number of bin pairs $H$, as illustrated in Fig.~\ref{fig:qcirc_one_step}.
The $U_{\text{add}}$ gate is called $H-1$ times because the increment of the value $1$ is not needed at the final division.
The $U_{\Delta t}$ operation is performed $M$ times inside $U_t$, as shown in Fig.~\ref{fig:qcirc_all_step}.

For the expected-value calculation, the cost of implementing $U_c$ is detailed in Appendix~\ref{appendix:quantum_algorithm_for_components}.
In summary, it is given by 
\begin{equation}
  T_\text{count}(U_c) = 1.15n_i^{\max}\log_2(n_i^{\max}/\epsilon_c),
\end{equation}
where $n_i^{\max}$ is the maximum number of droplets in the $i$-th bin defined above, and $\epsilon_c$ is the error tolerance of the $U_c$ operation.

Finally, the number of applications of $U_t$ and $U_c$ depends on the quantum amplitude estimation method. We use the iterative quantum amplitude estimation~\cite{Grinko2021-nk}, in which multiple rounds of circuits are executed. In each round, the Grover oracle is applied a chosen number of times, followed by a measurement. The Grover oracle contains one application of $A$ and one of its inverse $A^\dagger$, which encode and decode the quantum amplitude of interest. It is known that the total number of oracle applications summed over all rounds, which is denoted by $N_\text{oracle}$, can be bounded as
\begin{equation}
  \label{eq:iteration}
  N_\text{oracle} \lesssim  \frac{1.4}{\epsilon_\text{estimation}}\ln\left(\frac{2}{\delta}\log_2\left(\frac{\pi}{4\epsilon_\text{estimation}}\right)\right),
\end{equation}
where $\epsilon_\text{estimation}$ is the error tolerance of the amplitude estimation, defined below, and $\delta$ is the failure probability~\cite{Grinko2021-nk}.
The reported T-counts are the cost of a single Grover-oracle application, which contains both $A$ and $A^\dagger$, multiplied by $N_\text{oracle}$. The resulting cost scales as $1/\epsilon_\text{estimation}$, providing a quadratic improvement over direct sampling. Since these estimates are for a single bin, obtaining the expected droplet number in all $N$ bins multiplies the total gate count by $N$, while leaving the logical-qubit count unchanged.

In the gate-count estimates above, the following error sources are taken into account.
\begin{itemize}
  \item $\epsilon_\text{estimation}$ denotes the error in the quantum amplitude estimation, introduced in Eq.~\eqref{eq:iteration};
  \item $\epsilon_\text{calculation}$ denotes the error in the probability calculation, including errors caused by imperfect fixed-point arithmetic operations and the piecewise arcsine calculation (the tolerance $\epsilon_{\arcsin}$ introduced above is one contribution to it);
  \item $\epsilon_\text{rotation}$ denotes the error in the controlled rotation operations, which is the tolerance $\epsilon_{\sin}$ introduced above; and
  \item $\epsilon_c$ denotes the error in encoding the expected number of droplets into the quantum amplitude by the $U_c$ operation.
\end{itemize}
The total error is constrained by the cumulative effects of these errors, scaled by the number of times each operation is performed.
Therefore, its upper bound $\epsilon_{\max}$ can be calculated as
\iftwocolumnversion
\begin{multline}
  \epsilon_{\max} =  2N_\text{oracle}MH\left(\epsilon_\text{calculation} + \epsilon_\text{rotation}\right) \\
  + 2N_\text{oracle}\epsilon_c + \epsilon_\text{estimation}.
\end{multline}
\else
\begin{equation}
  \epsilon_{\max} =  2N_\text{oracle}MH\left(\epsilon_\text{calculation} + \epsilon_\text{rotation}\right) + 2N_\text{oracle}\epsilon_c + \epsilon_\text{estimation}.
\end{equation}
\fi
Note that this expression is not circular in $\epsilon_\text{estimation}$: $N_\text{oracle}$ is fixed beforehand by the chosen $\epsilon_\text{estimation}$ through Eq.~\eqref{eq:iteration}, and the expression above then evaluates $\epsilon_{\max}$ for that choice.

We note that $\epsilon_{\max}$ accounts for the errors of the quantum implementation but not for the error of the time discretization. The first-order accurate finite-difference scheme in Eq.~\eqref{eq:master_discretized} has a time-stepping error of $\mathcal{O}(\Delta t)$, which is controlled independently of $\epsilon_{\max}$ by choosing the number of time steps $M$. With $t_\text{final}=M\Delta t$ held fixed, increasing $M$ reduces this discretization error but raises both the T-count and the number of history qubits in proportion to $M$. Therefore, the number of time steps is a key parameter governing the trade-off between accuracy and cost.

Table~\ref{table:parameters} lists the parameters adopted in our resource estimation.
Five different cases are considered, with differences in $N$, $M$, and the maximum error $\epsilon_{\max}$.
The parameter $n_{\epsilon}$ fixes the round-off error of the fixed-point arithmetic, while $d_{\epsilon}$ and $S_{\epsilon}$ are fixed by $\epsilon_{\arcsin}$ through Table~\ref{table:piecewise_arcsin}. Both contribute to $\epsilon_\text{calculation}$, which is chosen carefully to keep the errors sufficiently small.

The results of our quantum resource estimates are summarized in Table~\ref{table:quantum_resources}.
Those estimates are for the calculation of the expected value of the number of droplets in the first bin of the final quantum state.
The totals in Table~\ref{table:quantum_resources} include state preparation (zero T gates), the full $M$-step time evolution $U_t$, and the $N_\text{oracle}$ Grover-oracle applications for measurement.

It is found that the required number of logical qubits ranges from $10^4$ to $10^5$, while the T-count is on the order of $10^{14}$ to $10^{16}$.
The T-count increases by approximately a factor of 160 for every tenfold increase in $N$.
When $M$ increases or the maximum error $\epsilon_{\max}$ decreases by a factor of 10, the T-count grows by roughly an order of magnitude in both cases.
The growth is somewhat larger for $\epsilon_{\max}$ because $n_{\epsilon}$, $\epsilon_\text{rotation}$, and $\epsilon_c$ are tightened simultaneously in case No.~5, so the increase is not due to the $1/\epsilon_\text{estimation}$ scaling of $N_\text{oracle}$ alone.
As shown in Table~\ref{table:quantum_resources}, the T-depth is consistently proportional to the T-count across all cases, with a ratio of approximately $0.7$--$0.8$, and therefore exhibits the same scaling behavior.
Notably, the number of logical qubits behaves differently: it increases by approximately a factor of 10 for a tenfold increase in $M$, but increases by at most a factor of 2 or remains constant for a tenfold increase in $N$ or a tenfold decrease in $\epsilon_{\max}$.

\begin{figure}[ht]
  \centering
  \includegraphics[width=\linewidth]{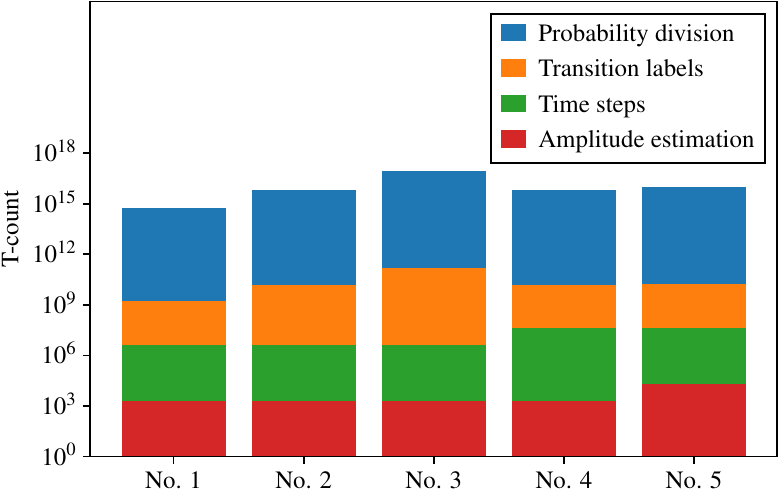}
  \caption{Stacked bar charts showing how many times each child subroutine is invoked by its parent, decomposing the T-counts by component.
  ``Probability division'' refers to the T-count for a single transition label of the probability division.
  ``Transition labels'' represents the number of transition labels, corresponding to the number of individual probability divisions invoked within $U_{\Delta t}$.
  ``Time steps'' denote the number of $U_{\Delta t}$ operations executed within $U_t$.
  ``Amplitude estimation'' reflects the number of $U_t$ operations invoked during a quantum amplitude estimation. The cost of the $U_c$ operation is omitted as its contribution to the total cost is negligible.}
  \label{fig:task_count}
\end{figure}

\begin{figure}[ht]
  \centering
  \includegraphics[width=\linewidth]{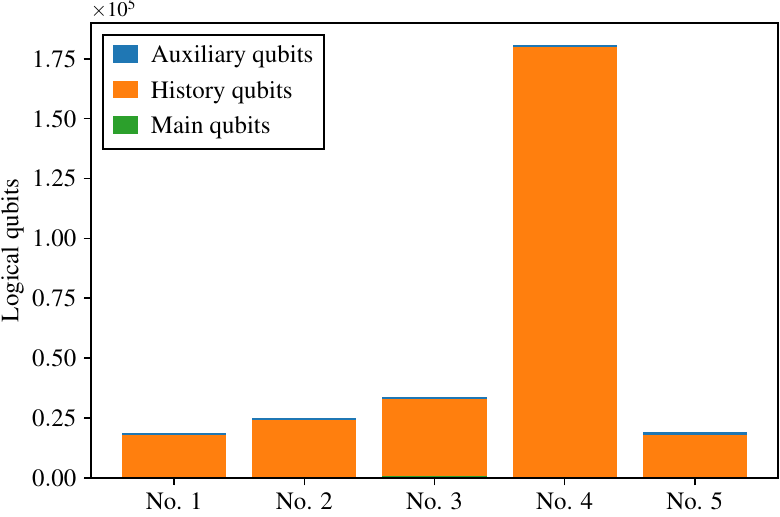}
  \caption{The detailed breakdown of the number of logical qubits.
  ``Auxiliary qubits'' are the sum of the numbers of logical qubits for $\mathcal{A},\mathcal{B},\mathcal{C}$, $\mathcal{D}$, and those used for quantum arithmetic.
  ``History qubits'' are used to encode the transition history.
  ``Main qubits'' are used to encode the droplet mass distribution.
  The numbers of auxiliary and main qubits are significantly smaller than the number of history qubits and barely visible in the figure.}
  \label{fig:qubits_count}
\end{figure}

To identify which components dominate the overall cost, we separately analyze the T-counts and the logical-qubit counts. For T gates, we quantify the contribution of each componentby counting how many times each child subroutine is invoked by its parent since our quantum circuit is built from a hierarchy of nested subroutines. For logical qubits, we decompose the total into the number of qubits allocated to each register role.
Both analyses are presented as stacked bar charts in Figs.~\ref{fig:task_count} and~\ref{fig:qubits_count}, respectively.
For T gates, the contributions are approximately even across all components in the baseline case (No.~1).
As shown by the comparison among Nos.~1--3, a growth in the total number of transition labels becomes the dominant factor as $N$ increases. Comparing No.~1 and No.~4, we find that an increase in $M$ makes the time-step contribution dominant. Comparing No.~1 and No.~5, we find that reducing the maximum error makes the contribution from the amplitude estimation dominant.
In contrast, for logical qubits, the primary cost contribution comes from the history register, whose number is proportional to $M$.

\begin{table}[ht]
  \centering
  \caption{\label{table:parameters} Parameters used in the estimation of quantum resources.
  $N$ is the number of mass bins, $M$ is the number of time steps, $n_{\epsilon}$ is the number of qubits for the piecewise arcsine calculation, $d_{\epsilon}$ and $S_{\epsilon}$ are the parameters of the piecewise arcsine calculation, $\epsilon_{\text{rotation}}$ is the error in the controlled rotation operation, $\epsilon_{\text{estimation}}$ is the error in the quantum amplitude estimation, and $\epsilon_c$ is the error in the $U_c$ operation.
  In addition, $\delta=0.01$ is used for the failure probability in the quantum amplitude estimation for all cases.
  The listed $d_{\epsilon}$ and $S_{\epsilon}$ correspond to the rows $\epsilon_{\arcsin}=10^{-12}$, $10^{-13}$, $10^{-15}$, $10^{-13}$, and $10^{-15}$ of Table~\ref{table:piecewise_arcsin} for cases No.~1--5, respectively. In all cases $\epsilon_\text{estimation}$ is set to $0.99\epsilon_{\max}$, so that the amplitude estimation dominates the error budget.}
  \begin{ruledtabular}
    \begin{tabular}{cccccccccc}
      case & $N$ & $M$ & $n_{\epsilon}$ & $d_{\epsilon}$ & $S_{\epsilon}$ & $\epsilon_{\text{rotation}}$ & $\epsilon_{\text{estimation}}$ & $\epsilon_c$ \\
      \hline
      No.~1 & 40  & 2000 & 42 & 5 & 15 & $10^{-13}$ & $9.9\times 10^{-3}$ & $10^{-8}$ \\
      No.~2 & 126 & 2000 & 46 & 6 & 12 & $10^{-14}$ & $9.9\times 10^{-3}$ & $10^{-8}$ \\
      No.~3 & 400 & 2000 & 49 & 8 & 10 & $10^{-15}$ & $9.9\times 10^{-3}$ & $10^{-8}$ \\
      No.~4 & 40 & 20000 & 46 & 6 & 12 & $10^{-14}$ & $9.9\times 10^{-3}$ & $10^{-9}$ \\
      No.~5 & 40 & 2000 & 49 & 8 & 10 & $10^{-15}$ & $9.9\times 10^{-4}$ & $10^{-10}$
    \end{tabular}
  \end{ruledtabular}
\end{table}

\begin{table}[ht]
  \centering
  \caption{\label{table:quantum_resources} Estimated quantum resources required for calculating the expected value of the number of droplets in the first bin.
  The resources include the maximum error $\epsilon_{\max}$, the T-count, the T-depth, and the number of logical qubits.
  T-depth is the number of layers of T gates when gates that can be applied in parallel are grouped together.}
  \begin{ruledtabular}
    \begin{tabular}{ccccc}
      case & $\epsilon_{\max}$ & T-count & T-depth & logical qubits \\
      \hline
      No.~1 & $1.0\times 10^{-2}$ &  $5.5\times 10^{14}$ &  $3.8\times 10^{14}$ &  $1.9\times 10^4$ \\
      No.~2 & $1.0\times 10^{-2}$ &  $6.9\times 10^{15}$ &  $5.0\times 10^{15}$ &  $2.5\times 10^4$ \\
      No.~3 & $1.0\times 10^{-2}$ &  $9.0\times 10^{16}$ &  $6.9\times 10^{16}$ &  $3.4\times 10^4$ \\
      No.~4 & $1.0\times 10^{-2}$ &  $6.9\times 10^{15}$ &  $5.0\times 10^{15}$ &  $1.8\times 10^5$ \\
      No.~5 & $1.0\times 10^{-3}$ &  $9.5\times 10^{15}$ &  $7.3\times 10^{15}$ &  $1.9\times 10^4$
    \end{tabular}
  \end{ruledtabular}
\end{table}

\section{\label{Discussion}Discussion}
In this collision-coalescence problem, the computational cost increases super-polynomially in classical calculations as the number of states grows~\cite{Alfonso2015-vq}.
For example, if the computation for $N=40$ requires one day, the same calculation for $N=126$ would take 250 years, and for $N=400$, it would take $5\times 10^{11}$ years, which is an impractical time frame.
In contrast, for fixed $M$ and $\epsilon_{\max}$, the T-count of the proposed quantum method scales as $\widetilde{\mathcal{O}}(N^2)$.
The improvement in the $N$-dependence arises because the transition operations act coherently on the encoded state without explicitly enumerating all possible mass distributions; QAE provides a separate quadratic improvement in the dependence on the target accuracy relative to direct sampling.

Here, the $\widetilde{\mathcal{O}}(N^2)$ scaling is with respect to the number of mass bins $N$, with $M$ and $\epsilon_{\max}$ fixed. The $\widetilde{\mathcal{O}}$ notation denotes the soft-O notation, which suppresses factors polylogarithmic in $N$. The quadratic factor $N^2$ originates from the number of transition labels $H\sim N^2/4$ in Eq.~\eqref{eq:H_count}. Each transition requires a single probability division. The cost of one division depends on $N$ only through the register sizes, which grow logarithmically as $q_h=\lceil\log_2(H+1)\rceil=\mathcal{O}(\log N)$. The fixed-point precision $n_\epsilon$ that is associated with the arithmetic cost is set to satisfy the target accuracy and grows at most logarithmically with $N$ since holding $\epsilon_{\max}$ fixed requires the per-operation tolerance to be tightened in proportion to $H\sim N^2/4$. These polylogarithmic $N$-factors are precisely absorbed in the soft-O notation. The full cost also grows linearly with $M$ and with $1/\epsilon_{\max}$ through $N_\text{oracle}$ because $\epsilon_\text{estimation}$ is set proportional to $\epsilon_{\max}$. The total T-count per bin therefore scales as $\widetilde{\mathcal{O}}(N^2 M/\epsilon_{\max})$, while reading out all $N$ bins multiplies this count by $N$. Therefore, the advantage over the classical computation lies in replacing the super-polynomial growth in $N$ (originating from the rapidly increasing number of possible states in Eq.~\eqref{eq:state_number}) by a polynomial one, rather than in removing the dependence on $M$ or $\epsilon_{\max}$.

A further clarification concerns the classical baseline used in this comparison. We compare the proposed quantum method with a direct state-space solution of the master equation, in which the probabilities of all possible states are explicitly evolved, rather than with sampling-based approaches such as Monte Carlo or the super-droplet method. The scaling comparison above therefore concerns the cost of avoiding this explicit enumeration of the master-equation state space~\cite{Alfonso2015-vq,Dziekan2017-wb}.

To achieve the speedup, we extended the quantum algorithm originally developed for financial applications~\cite{Wang2024-me} by introducing a key modification to the state encoding.
In our quantum algorithm, the qubits are separated into two parts: one encodes the system state and the other encodes the transition history.
In a naive encoding, one would track all $R(N)$ possible mass distributions in Eq.~\eqref{eq:state_number} at every time step. Since $R(N)$ grows super-polynomially with $N$, this requires $\mathcal{O}(\log R(N))=\mathcal{O}(\sqrt{N})$ qubits per step. Thus, the cost is polynomial in $N$. In contrast, our algorithm records only the transition history $\bm{h}=(h_1,h_2,\ldots,h_M)$ in Eq.~\eqref{eq:transition_history}, where each $h_m$ takes one of $H\sim N^2/4$ transition labels in Eq.~\eqref{eq:H_count}. Because $H$ is polynomial in $N$, the history register requires only $\lceil\log_2(H+1)\rceil = \mathcal{O}(\log N)$ qubits per step. As a result, the cost is polylogarithmic in $N$.

We emphasize that this memory-efficient encoding is not specific to the problem of cloud droplet growth, but is potentially applicable to other problems that involve the time evolution of spectral probability distributions.
For instance, our quantum algorithm may be useful in solving the probability distribution of turbulent kinetic energy spectra using a master equation with a collision kernel, or in computing the time evolution of chemical reaction networks with analogous reaction-channel structures~\cite{McQuarrie1967-rz,Gillespie1992-uf}.
Exploring such applications is an important direction for future research.

It would be useful to note the limitation of the proposed quantum algorithm.
One major challenge is its high base computational cost.
The T-count is typically at most $10^{14}$ in many quantum algorithms that are deemed promising, such as those for quantum chemistry~\cite{von-Burg2021-ze,Rubin2023-wi,Lee2021-mj}.
While our algorithm significantly suppresses the growth in computational cost compared to the classical methods, it still requires substantial resources for the probability division and for handling time steps.
Notably, this challenge is not unique to this study but is shared by the quantum finance areas~\cite{Wang2024-me}.
Substantial progress is highly anticipated for these topics to make problems tractable.

To overcome this challenge, several strategies may be possible.
One approach is to modify the problem settings to reduce computational demands.
For instance, a piecewise approach could be employed to limit the number of time steps, rather than solving the system continuously across all time steps.

Another strategy is to utilize size-based distribution functions, as in spectral bin methods, to constrain the number of bins.
These strategies represent potential future research directions, aiming at reducing the computational cost of the quantum algorithm.

Despite this limitation, our results suggest that quantum computing has significant, yet still latent, potential to solve problems in atmospheric science, particularly in the cases where quantities of interest are expectations over probability distributions.
This study demonstrated its potential only on the computation of the expected values at the final time, but it also allows for extracting characteristic features from the time-series data since the calculated quantum state preserves all temporal information.
Such analysis would be explored in future work.

\section{\label{Conclusion}Conclusion}
We have developed a quantum algorithm to simulate the collision-coalescence process of cloud droplets.
The computation is based on the master equation that governs the time evolution of the probability distribution of droplet masses.
By encoding this probability distribution into the quantum amplitudes, our algorithm can represent super-polynomially large state spaces efficiently in terms of qubit counts.

Unlike the conventional approaches in quantum finance, which explicitly encode the full distribution (e.g., mass distributions) at each time step, our method encodes only the history of transition events.
This allows us to avoid storing all intermediate distributions, and to significantly reduce the number of qubits required for representing the time evolution.
The time evolution is performed via the probability-division operations, which can be executed in parallel for the same transition labels.
As a result, the computational cost scales only with the number of distinct transition labels.
We have constructed the quantum circuit for the probability divisions using the quantum arithmetic operations and the controlled rotation gates.

The algorithm is designed to estimate the expected number of droplets in each mass bin at the final time.
Instead of measuring the full quantum state, the quantum amplitude estimation is used to directly estimate the amplitudes corresponding to the expected values.
This reduces the number of repetitions required for readout.

We have estimated the quantum resources of the new algorithm by analyzing the logical-qubit count and the T-count.
The T-count scales as $\widetilde{\mathcal{O}}(N^2)$ per mass bin with respect to the number of mass bins $N$, for a fixed number of time steps and target accuracy, which is significantly smaller than the super-polynomial cost of the classical master-equation simulations.
However, the computational cost becomes high if the number of time steps increases.
This point should be addressed in future work.

Our results demonstrate that quantum computing has the potential to significantly reduce the computational cost of the simulation of the collision-coalescence process of cloud droplets.
Thus, we conclude that the collision-coalescence process is one of the promising targets of quantum computing in atmospheric science.

\begin{acknowledgments}
  We are grateful to Prof.~Shinichiro Shima, Dr.~Keita Kanno, and Dr.~Yasunori Lee for helpful discussions.
  This work was supported by JST SPRING, Grant Number JPMJSP2108 and International Graduate Program for Excellence in Earth-Space Science (IGPEES), a World-leading Innovative Graduate Study (WINGS) Program, the University of Tokyo.
\end{acknowledgments}

\appendix

\section[Fixed-point quantum arithmetic operations]{Fixed-point quantum arithmetic operations}
\label{appendix:fixed_point}
The fixed-point quantum arithmetic operations are used to calculate the transition probabilities.
In this section, we explain the costs of implementing the fixed-point quantum arithmetic operations. 

In our calculations, all values are positive. Also, as we show later, the values are either real numbers in the range of $[0,\pi/2]$ or integers.
The real values are represented by the fixed-point numbers with the precision of $n$ bits as 
\begin{equation}
  \label{eq:real_in_0_pi/2}
  \alpha = \alpha_{n-1}2^{0} + \alpha_{n-2}2^{-1} + \cdots + \alpha_{0}2^{-n+1},
\end{equation}
where $\alpha_{i}\in \{0,1\}$ for each $i$ and we can rewrite $\alpha$ as $\alpha_{n-1}.\alpha_{n-2}\cdots\alpha_{0}$ in the binary representation.
The integer values are represented as
\begin{equation}
  \label{eq:integer}
  \alpha = \alpha_{n-1}2^{n-1} + \alpha_{n-2}2^{n-2} + \cdots + \alpha_{0}2^{0},
\end{equation}
and $\alpha_{n-1}\alpha_{n-2}\cdots\alpha_{0}$ in the binary representation.
Throughout this appendix, $n$ denotes the generic bit precision of an arithmetic primitive. In our algorithm, the primitives acting on the registers $\mathcal{A}$ and $\mathcal{B}$ are instantiated with $n=n_{\epsilon}$.

Most of the calculations used in our algorithm follow those used in~\cite{Wang2024-me}, and this is summarized in their Appendix. We summarize the results of their resource estimation for these calculations in Table~\ref{table:fixed_point_arithmetics}.
In addition, we explain briefly the operation that is different from the one in~\cite{Wang2024-me} in the following.

\subsection{Comparison}
The comparison operation is expressed as
\begin{equation}
  \text{COMP}_n : \ket{\alpha}\ket{\beta}\ket{0} \mapsto \begin{cases}
    \ket{\alpha}\ket{\beta}\ket{1} & \text{if } \alpha \geq \beta, \\
    \ket{\alpha}\ket{\beta}\ket{0} & \text{otherwise}.
  \end{cases}
\end{equation}
This operation is part of the COMP\_CONST operation in \cite{Wang2024-me}. The resulting circuit has
\begin{align}
  \nonumber
  T_\text{count}(\text{COMP}_n) &= T_\text{count}(\text{SUB}_{n-1}) + T_\text{count}(\text{ADD}_{n-1}) \\
  &= 8n-16
\end{align}
T gates. T-depth is $4n-8$ and the number of auxiliary qubits is $2n-1$.
We also use the comparison against a classical constant, which is denoted by $\text{COMP\_CONST}_n$ and is obtained by replacing one of the two inputs of $\text{COMP}_n$ with a constant. Its cost is bounded by that of $\text{COMP}_n$, and we adopt the T-count of $\text{COMP}_n$ for it in our estimation.

\subsection{Multiplication}
The fixed-point numbers we used here are expressed as either Eq.~\eqref{eq:real_in_0_pi/2} or Eq.~\eqref{eq:integer}.
As we show later, we only use the multiplication whose result is either an integer or lies between 0 and 1.
We use standard shift-and-add strategy, which is also used in~\cite{Wang2024-me}.
The multiplication between two integers $\alpha$ and $\beta$ is expressed as
\begin{equation}
  \text{MUL\_INT}_{n,m} : \ket{\alpha}\ket{\beta}\ket{0} \mapsto \ket{\alpha}\ket{\beta}\ket{\alpha\beta},
\end{equation}
where $n$ and $m$ are the number of bits of $\alpha$ and $\beta$, respectively, and $n\leq m$.
The required T-count is
\begin{align}
  \nonumber
  T_\text{count}(\text{MUL\_INT}_{n,m}) &= \sum_{i=0}^{n-1}T_\text{count}(\text{c-ADD}_{m-i}) \\
  \nonumber
  &= \sum_{i=0}^{n-1}(8(m-i)-4) \\
  &= 8nm-4n^2,
\end{align}
T-depth is $4nm-2n^2$ and the number of auxiliary qubits is $2n-1$.

We also use the multiplication between two real numbers $\alpha$ and $\beta$ that are in the range of $[0,1]$ and both of which have bit length $n$. This operation is expressed as 
\begin{equation}
  \text{MUL\_UI}_n : \ket{\alpha}\ket{\beta}\ket{0} \mapsto \ket{\alpha}\ket{\beta}\ket{\overline{\alpha\beta}},
\end{equation}
where $\overline{\alpha\beta}$ is the $n$ bit fixed-point representation of $\alpha \beta$. Here and throughout, $\overline{x}$ denotes the $n$-bit fixed-point representation of a real number $x$.
The required T-count is
\begin{align}
  \nonumber
  T_\text{count}(\text{MUL\_UI}_n) &= \sum_{i=0}^{n-1}T_\text{count}(\text{c-ADD}_{n-i}) \\
  \nonumber
  &= \sum_{i=0}^{n-1}(8(n-i)-4) \\
  &= 4n^2,
\end{align}
T-depth is $2n^2$ and the number of auxiliary qubits is $2n-1$.

When calculating the multiplication between an integer and a constant real number, we use the operation 
\begin{equation}
  \text{MUL\_CONST\_INT\_UI}_{n,m} : \ket{\alpha}\ket{0} \mapsto \ket{\alpha}\ket{\overline{\alpha\beta}}, 
\end{equation}
where $n$ is the number of bits of the integer and $m$ is the number of bits of the constant real number.
The required T-count is
\begin{align}
  \nonumber
  T_\text{count}&(\text{MUL\_CONST\_INT\_UI}_{n,m}) 
  \\ & \nonumber
  = (m-n)T_\text{count}(\text{ADD}_{n}) +  \sum_{i=0}^{n-1}T_\text{count}(\text{ADD}_{n-i}) \\
  \nonumber
  &= (m-n)(4n-4) + \sum_{i=0}^{n-1}(4(n-i)-4) \\
  &= 4nm-2n^2-4m+2n.
\end{align}
T-depth is $2nm-n^2-2m+n$ and the number of auxiliary qubits is $n-1$.

\subsection[Square root of a real number between 0 and 1]{Square root of a real number between 0 and 1}
The square root is described in~\cite{Wang2024-me}, but we modified the operation to be suitable for our calculations with decimals.
The square root of the fixed-point number $\alpha$ is calculated as
\begin{equation}
  \text{SQRT}_n : \ket{\alpha}\ket{0} \mapsto \ket{\alpha}\ket{\overline{\sqrt{\alpha}}}.
\end{equation}
In our algorithm, $\alpha \in [0,1]$. Reference~\cite{Wang2024-me} gives an operation in which the number of digits in the result is halved.
We modified this so that the operation is conducted until the last bit is reached. The resulting circuit has
\begin{align}
  \nonumber
  T_\text{count}(\text{SQRT}_n) &= 2\sum_{i=1}^{n-1}T_\text{count}(\text{ADD}_{2i})
  \iftwocolumnversion \nonumber \\ &\quad \fi
  + 2T_\text{count}(\text{c-ADD}_{2(n-1)}) \\
  \nonumber
  &= 2\sum_{i=1}^{n-1}(8i-4) + 2(16(n-1)-4) \\
  &= 8n^2 + 16n - 32
\end{align}
T gates. T-depth is $4n^2+8n-16$ and the number of auxiliary qubits is $6n$.

\subsection[Division of a real number between 0 and 1]{Division of a real number between 0 and 1}
The quantum division is expressed as 
\begin{equation}
  \label{eq:DIV}
  \text{DIV}_n : \ket{\alpha}\ket{\beta}\ket{0} \mapsto \ket{\alpha}\ket{\beta}\ket{\overline{\alpha/\beta}}.
\end{equation}
In our algorithm, $\alpha,\beta, \alpha/\beta \in [0,1]$. Reference~\cite{Orts2024-xz} proposed the so-called ``long division'' algorithm.
This is an algorithm for integer division, but we can apply it to the division of real numbers between 0 and 1 and change it to calculate until the last bit is reached.
Also, by using the do-copy-undo trick, we change it to the clean version described as Eq.~\eqref{eq:DIV}.
The resulting circuit has
\begin{align}
  \nonumber
  T_\text{count}&(\text{DIV}_n) 
  \\ \nonumber &
  = 2\sum_{i=0}^{n-1}\left[T_\text{count}(\text{COMP}_{n-i}) + 2T_\text{count}(\text{c-SUB}_{n-i}) \right]\\
  \nonumber
  &= 2\sum_{i=0}^{n-1}\left[8(n-i)-16 + 2(8(n-i)-4) \right] \\
  &= 24n^2 - 24n
\end{align}
T gates. T-depth is $12n^2-12n$ and the number of auxiliary qubits is $2n-1$.

\subsection{Arcsine}
The calculation of the arcsine is approximated by the piecewise polynomial function.
This is expressed as
\begin{equation}
  \text{ARCSIN}_{n,\epsilon_{\arcsin}} : \ket{\alpha}\ket{0} \mapsto \ket{\alpha}\ket{\overline{\text{pp}_{\arcsin}(\alpha)}}.
\end{equation}
In our algorithm, $\alpha \in [0,0.5]$ and $\overline{\text{pp}_{\arcsin}(\alpha)} \in [0,1]$.
We apply the piecewise polynomial function described in ~\cite{Wang2024-me} to the arcsine calculation and our operation corresponds to $p=1$ case.
The resulting circuit has
\begin{align}
  \nonumber
  T_\text{count}&(\text{ARCSIN}_{n,\epsilon_{\arcsin}}) 
  \\ \nonumber &
  = 4S_{\epsilon}T_\text{count}(\text{COMP\_CONST}_n) 
  \\ \nonumber &
  + 2d_{\epsilon}[T_\text{count}(\text{MUL\_UI}_n)+T_\text{count}(\text{ADD}_n)] \\
  \nonumber
  &+ 4d_{\epsilon}S_{\epsilon}T_\text{count}(\text{Toffoli}_{\lceil \log_2(S_{\epsilon}) \rceil+1}) \\
  =& 4S_{\epsilon}(8n-16) + 2d_{\epsilon}(4n^2+4n-4) 
  \\ \nonumber &
  + 16d_{\epsilon}S_{\epsilon}(\lceil \log_2(S_{\epsilon}) \rceil-1)
\end{align}
T gates.
$S_{\epsilon}$ and $d_{\epsilon}$ are the number of pieces and the degree of the polynomial function corresponding to the precision $\epsilon_{\arcsin}$, respectively; the subscript $\epsilon$ of these two symbols refers to this tolerance. The number of pieces is denoted by $M_{\epsilon}$ in~\cite{Wang2024-me}; we use $S_{\epsilon}$ here to avoid confusion with the number of time steps $M$.
T-depth is $4d_{\epsilon}\max (2n^2, S_{\epsilon}(\lceil \log_2(S_{\epsilon}) \rceil-1)) + 16S_{\epsilon}(n-2)+4d_{\epsilon}(n-1)$ and the number of auxiliary qubits is $(d_{\epsilon}+4)n+2\lceil \log_2(S_{\epsilon}) \rceil$.

In order to determine $S_{\epsilon}$ and $d_{\epsilon}$ for given $\epsilon_{\arcsin}$, we take a strategy of finding the minimum $S_{\epsilon}$ for given $d_{\epsilon}$.
Let $\Omega = [0,0.5]$.
We conduct the following steps to determine the minimum $S_{\epsilon}$:
\begin{enumerate}
  \item Set $\Omega_1 = [0,0.5]$. Calculate the Chebyshev polynomial of degree $d_{\epsilon}$ that approximates the arcsine function on $\Omega_1$.
  \item Calculate the $L_{\infty}$-error of the Chebyshev polynomial on $\Omega_1$.
  \item If the error is less than $\epsilon_{\arcsin}$, the minimum $S_{\epsilon}$ is 1. If not, set $\Omega_1 = [0,(0+0.5)/2]$.
  \item Repeat the steps 1--3 until the error is less than $\epsilon_{\arcsin}$, and determine the first subdomain $\Omega_1=[0,a_1]$ that satisfies the condition.
  \item Set $\Omega_2 = [a_1,0.5]$ and repeat the steps 1--4 to determine the second subdomain $\Omega_2=[a_2,a_3]$.
  \item By repeating the steps 1--5 until $a_i$ reaches 0.5, the minimum $S_{\epsilon}$ can be determined.
\end{enumerate}
The result of this calculation is summarized in Table~\ref{table:piecewise_arcsin}.
In our actual calculations, we choose $d_{\epsilon}$ and $S_{\epsilon}$ that minimize the T-count from this table.

\begin{table*}
  \caption{\label{table:fixed_point_arithmetics}The T-count and the number of auxiliary qubits required for each fixed-point quantum arithmetic operation used in our cost estimation.}
    \begin{ruledtabular}
      \begin{tabular}{ccccc}
        Operation & T-count & T-depth & auxiliary qubits & techniques \\
        \hline
        $\text{Toffoli}_n$ & $4n-8$ & $n-2$ & $n-1$ & \cite{Wang2024-me}, \cite{Jones2013-vi} \\
        $\text{ADD}_n$ & $4n-4$ & $2n-2$ & $n-1$ & \cite{Wang2024-me}, \cite{Gidney2018-tj} \\
        $\text{SUB}_n$ & $4n-4$ & $2n-2$ & $n-1$ & \cite{Wang2024-me}, \cite{Gidney2018-tj} \\
        $\text{c-ADD}_n$ & $8n-4$ & $4n-2$ & $2n-1$ & \cite{Wang2024-me}, \cite{Gidney2018-tj} \\
        $\text{c-SUB}_n$ & $8n-4$ & $4n-2$ & $2n-1$ & \cite{Wang2024-me}, \cite{Gidney2018-tj} \\
        $\text{ADD\_CONST}_n$ & $4n-8$ & $2n-4$ & $2n-2$ & \cite{Wang2024-me}, \cite{Gidney2018-tj} \\
        $\text{COMP}_n$ & $8n-16$ & $4n-8$ & $2n-1$ & \cite{Wang2024-me}, \cite{Gidney2018-tj} \\
        $\text{MUL\_INT}_{n,m}$ & $8nm-4n^2$ & $4nm-2n^2$ & $2n-1$ & \cite{Wang2024-me}, \cite{Gidney2018-tj} \\
        $\text{MUL\_UI}_{n}$ & $4n^2$ & $2n^2$ & $2n-1$ & \cite{Wang2024-me}, \cite{Gidney2018-tj} \\
        \shortstack{$\text{MUL\_CONST\_INT\_UI}_{n,m}$\\{}} & \shortstack{$4nm-2n^2$ \\ $-4m+2n$} & \shortstack{$2nm-n^2$ \\ $-2m+n$} & \shortstack{$n-1$\\{}} & \shortstack{\cite{Wang2024-me}, \cite{Gidney2018-tj}\\{}} \\
        $\text{SQRT}_n$ & $8n^2+16n-32$ & $4n^2+8n-16$ & $6n$ & \cite{Wang2024-me}, \cite{Gidney2018-tj}, \cite{Munoz-Coreas2018-yg} \\
        $\text{DIV}_n$ & $24n^2-24n$ & $12n^2-12n$ & $2n-1$ & \cite{Orts2024-xz} \\
        \shortstack{$\text{ARCSIN}_{n,\epsilon_{\arcsin}}$\\{}} & \shortstack{$32S_{\epsilon}(n-2)$ \\ $+ 8d_{\epsilon}(n^2+n-1)$ \\ $+ 16d_{\epsilon}S_{\epsilon}$ \\ $(\lceil \log_2(S_{\epsilon}) \rceil-1)$} & \shortstack{$4d_{\epsilon}\max (2n^2, S_{\epsilon}$ \\ $(\lceil \log_2(S_{\epsilon}) \rceil-1))$ \\ $+ 16S_{\epsilon}(n-2)$ \\$+4d_{\epsilon}(n-1)$} & \shortstack{$(d_{\epsilon}+4)n$ \\ $+ 2\lceil \log_2(S_{\epsilon}) \rceil$} & \shortstack{{}\\ \cite{Haner2018-jy}, \cite{Wang2024-me}, \\ \cite{Jones2013-vi}, \cite{Gidney2018-tj}, \\ \cite{Chakrabarti2021-qo}} \\
      \end{tabular}
    \end{ruledtabular}
\end{table*}

\begin{table}
  \caption{\label{table:piecewise_arcsin}The minimum $S_{\epsilon}$ and $d_{\epsilon}$ for given $\epsilon_{\arcsin}$ in the piecewise arcsine calculation.}
  \begin{ruledtabular}
    \begin{tabular}{ccc}
      $\epsilon_{\arcsin}$ & $d_{\epsilon}$ & $S_{\epsilon}$ \\
      \hline
      $10^{-12}$ & 4 & 43 \\
      & 5 & 15 \\
      & 6 & 9 \\
      $10^{-13}$ & 5 & 25 \\
      & 6 & 12 \\
      & 7 & 7 \\
      $10^{-14}$ & 5 & 35 \\
      & 6 & 18 \\
      & 7 & 10 \\
      & 8 & 7 \\
      $10^{-15}$ & 6 & 27 \\
      & 7 & 13 \\
      & 8 & 10 \\
      & 9 & 11 \\
    \end{tabular}
  \end{ruledtabular}
\end{table}

\section{Quantum algorithm for operators $U_P$, $U_{\sin}$, $U_Q$, $U_{\text{add}}$, $U_R$, $U_{\text{shift}}$, and $U_c$}
\label{appendix:quantum_algorithm_for_components}
\subsection{Quantum algorithm for $U_P$}
We utilize fixed-point quantum arithmetic operations to calculate transition probabilities. The operations are summarized in the Appendix of~\cite{Wang2024-me}.

As written in Eq.~\eqref{eq:U_P}, the $U_P$ gate calculates the piecewise arcsine of the square root of a modified transition probability.
The quantum circuit for $U_P$ is composed of the following operations:
\begin{enumerate}
  \item Copy the number of droplets in the $i(h)$-th and $j(h)$-th bins to auxiliary qubits $\mathcal{C}_1$ and $\mathcal{C}_2$.:
  \iftwocolumnversion
  \begin{multline}
    \ket{n_{i(h)}}_{\mathcal{N}_{i(h)}}\ket{n_{j(h)}}_{\mathcal{N}_{j(h)}}\ket{0}_{\mathcal{C}_1}\ket{0}_{\mathcal{C}_2}
    \\
    \mapsto \ket{n_{i(h)}}_{\mathcal{N}_{i(h)}}\ket{n_{j(h)}}_{\mathcal{N}_{j(h)}}\ket{n_{i(h)}}_{\mathcal{C}_1}\ket{n_{j(h)}}_{\mathcal{C}_2}.
  \end{multline}
  \else
  \begin{equation}
    \ket{n_{i(h)}}_{\mathcal{N}_{i(h)}}\ket{n_{j(h)}}_{\mathcal{N}_{j(h)}}\ket{0}_{\mathcal{C}_1}\ket{0}_{\mathcal{C}_2}
    \mapsto \ket{n_{i(h)}}_{\mathcal{N}_{i(h)}}\ket{n_{j(h)}}_{\mathcal{N}_{j(h)}}\ket{n_{i(h)}}_{\mathcal{C}_1}\ket{n_{j(h)}}_{\mathcal{C}_2}.
  \end{equation}
  \fi
  $\mathcal{C}_1$ and $\mathcal{C}_2$ are both $q_1$ qubits. In this step, CNOT gates are only used.
  \item Compute multiplication of $n_{i(h)}$ and $n_{j(h)}$ and store the result in auxiliary qubits $\mathcal{C}_3$:
  \iftwocolumnversion
  \begin{multline}
    \ket{n_{i(h)}}_{\mathcal{C}_1}\ket{n_{j(h)}}_{\mathcal{C}_2}\ket{0}_{\mathcal{C}_3}
    \\
    \mapsto \ket{n_{i(h)}}_{\mathcal{C}_1}\ket{n_{j(h)}}_{\mathcal{C}_2}\ket{n_{i(h)}n_{j(h)}}_{\mathcal{C}_3}.
  \end{multline}
  \else
  \begin{equation}
    \ket{n_{i(h)}}_{\mathcal{C}_1}\ket{n_{j(h)}}_{\mathcal{C}_2}\ket{0}_{\mathcal{C}_3}
    \mapsto \ket{n_{i(h)}}_{\mathcal{C}_1}\ket{n_{j(h)}}_{\mathcal{C}_2}\ket{n_{i(h)}n_{j(h)}}_{\mathcal{C}_3}.
  \end{equation}
  \fi
  $\mathcal{C}_3$ is $2q_1$ qubits. In this step, $\text{MUL\_INT}_{q_1,q_1}$ is used.
  \item Compute the multiplication of $n_{i(h)}n_{j(h)}$ and $K(i(h),j(h))\Delta t$ and store the result $r_h=n_{i(h)}n_{j(h)}K(i(h),j(h))\Delta t$ in auxiliary qubits $\mathcal{C}_4$:
  \begin{equation}
    \ket{n_{i(h)}n_{j(h)}}_{\mathcal{C}_3}\ket{0}_{\mathcal{C}_4} \mapsto \ket{n_{i(h)}n_{j(h)}}_{\mathcal{C}_3}\ket{r_h}_{\mathcal{C}_4}.
  \end{equation}
  $\mathcal{C}_4$ is $n_{\epsilon}$ qubits, where $n_{\epsilon}$ is the number of bits to express the transition probability. In this step, $\text{MUL\_CONST\_INT\_UI}_{2q_1,n_{\epsilon}}$ is used.
  \item Compare $r_h$ and $s_{h+1}/4$ and store the result in auxiliary qubits $\mathcal{C}_5$:
  \begin{equation}
    \ket{s_{h+1}}_{\mathcal{A}}\ket{r_h}_{\mathcal{C}_4}\ket{0}_{\mathcal{C}_5} \mapsto \ket{s_{h+1}}_{\mathcal{A}}\ket{r_h}_{\mathcal{C}_4}\ket{z}_{\mathcal{C}_5},
  \end{equation}
  where $z=1$ if $r_h \geq s_{h+1}/4$ and $z=0$ otherwise. $\mathcal{C}_5$ is $1$ qubit.
  In this step, $\text{COMP}_{n_{\epsilon}}$ is used. This step is needed to guarantee that the input to the arcsine function, which is used in step 9, does not exceed $1/4$.
  \item Based on the result of the comparison, store $\ket{w}$ in auxiliary qubits $\mathcal{C}_6$:
  \begin{equation}
    \ket{z}_{\mathcal{C}_5}\ket{0}_{\mathcal{C}_6} \mapsto \ket{z}_{\mathcal{C}_5}\ket{w}_{\mathcal{C}_6},
  \end{equation}
  where $w=s_{h+1}-r_h$ if $z=1$ and $w=r_h$ otherwise. $\mathcal{C}_6$ is $n_{\epsilon}$ qubits.
  In this step, $\text{c-SUB}_{n_{\epsilon}}$ is used.
  \item Compute the square root of $w$ and store the result in auxiliary qubits $\mathcal{C}_7$:
  \begin{equation}
    \ket{w}_{\mathcal{C}_6}\ket{0}_{\mathcal{C}_7} \mapsto \ket{w}_{\mathcal{C}_6}\ket{\sqrt{w}}_{\mathcal{C}_7}.
  \end{equation}
  $\mathcal{C}_7$ is $n_{\epsilon}$ qubits.
  In this step, $\text{SQRT}_{n_{\epsilon}}$ is used.
  \item Compute the square root of $s_{h+1}$ and store the result in auxiliary qubits $\mathcal{C}_8$: 
  \begin{equation}
    \ket{s_{h+1}}_{\mathcal{A}}\ket{0}_{\mathcal{C}_8} \mapsto \ket{s_{h+1}}_{\mathcal{A}}\ket{\sqrt{s_{h+1}}}_{\mathcal{C}_8}.
  \end{equation}
  $\mathcal{C}_8$ is $n_{\epsilon}$ qubits.
  In this step, $\text{SQRT}_{n_{\epsilon}}$ is used.
  \item Compute the division of $\sqrt{w}$ and $\sqrt{s_{h+1}}$ and store the result in auxiliary qubits $\mathcal{C}_9$:
  \iftwocolumnversion
  \begin{multline}
    \ket{\sqrt{w}}_{\mathcal{C}_7}\ket{\sqrt{s_{h+1}}}_{\mathcal{C}_8}\ket{0}_{\mathcal{C}_9} \\
    \mapsto \ket{\sqrt{w}}_{\mathcal{C}_7}\ket{\sqrt{s_{h+1}}}_{\mathcal{C}_8}\ket{\sqrt{w/s_{h+1}}}_{\mathcal{C}_9}.
  \end{multline}
  \else
  \begin{equation}
    \ket{\sqrt{w}}_{\mathcal{C}_7}\ket{\sqrt{s_{h+1}}}_{\mathcal{C}_8}\ket{0}_{\mathcal{C}_9} \mapsto \ket{\sqrt{w}}_{\mathcal{C}_7}\ket{\sqrt{s_{h+1}}}_{\mathcal{C}_8}\ket{\sqrt{w/s_{h+1}}}_{\mathcal{C}_9}.
  \end{equation}
  \fi
  $\mathcal{C}_9$ is $n_{\epsilon}$ qubits.
  In this step, $\text{DIV}_{n_{\epsilon}}$ is used.
  \item Compute the piecewise arcsine of $\sqrt{w/s_{h+1}}$ and store the result in auxiliary qubits $\mathcal{B}$:
  \begin{equation}
    \ket{0}_{\mathcal{B}}\ket{\sqrt{w/s_{h+1}}}_{\mathcal{C}_9} \mapsto \ket{\text{pp}_{\arcsin}{\sqrt{w/s_{h+1}}}}_{\mathcal{B}}\ket{\sqrt{w/s_{h+1}}}_{\mathcal{C}_9}.
  \end{equation}
  In this step, $\text{ARCSIN}_{n_{\epsilon},\epsilon_{\arcsin}}$ is used.
  \item Finally, set the state of the auxiliary qubits $\mathcal{B}$ to $\ket{\pi/2-\text{pp}_{\arcsin}{\sqrt{w/s_{h+1}}}}$ if $z=1$ and $\ket{\text{pp}_{\arcsin}{\sqrt{w/s_{h+1}}}}$ otherwise, which corresponds to the $\arcsin{\sqrt{r_h'}}$:
  \begin{equation}
    \ket{\text{pp}_{\arcsin}{\sqrt{w/s_{h+1}}}}_{\mathcal{B}}\ket{z}_{\mathcal{C}_5} \mapsto \ket{\text{pp}_{\arcsin}{\sqrt{r_h'}}}_{\mathcal{B}}\ket{z}_{\mathcal{C}_5}.
  \end{equation}
  In this step, $\text{c-SUB}_{n_{\epsilon}}$ and CNOT gates are used.
\end{enumerate}
In total, the $U_P$ gate requires 1 $\text{MUL\_INT}_{q_1,q_1}$, 1 $\text{MUL\_CONST\_INT\_UI}_{2q_1,n_{\epsilon}}$, 1 $\text{COMP}_{n_{\epsilon}}$, 2 $\text{c-SUB}_{n_{\epsilon}}$, 2 $\text{SQRT}_{n_{\epsilon}}$, 1 $\text{DIV}_{n_{\epsilon}}$, and 1 $\text{ARCSIN}_{n_{\epsilon},\epsilon_{\arcsin}}$. The T-count required for each operation is summarized in Table~\ref{table:fixed_point_arithmetics}.
The $U_P$ gate requires $\mathcal{C}_1$ to $\mathcal{C}_9$ auxiliary qubits and also requires additional auxiliary qubits to perform the quantum arithmetic operations. The number of auxiliary qubits required for each operation is summarized in Table~\ref{table:fixed_point_arithmetics}.

\subsection{Quantum algorithm for $U_{\sin}$}
By using $\mathcal{B}$ as a control qubit, the $U_{\sin}$ gate reflects the transition probability into the quantum amplitude when the transition history state is 0.
We slightly modify the method introduced in Appendix B of ~\cite{Wang2024-me} in order to operate $U_{\sin}$ only when the transition history state is 0. 
In what follows, $\mathcal{H}_m[k]$ denotes the $k$-th qubit of the register $\mathcal{H}_m$, where $k=0$ corresponds to the least significant bit, and $\mathcal{H}_m[k_2:k_1]$ denotes the set of qubits from $\mathcal{H}_m[k_1]$ to $\mathcal{H}_m[k_2]$. The same notation is used for the register $\mathcal{B}$.
What we want to do is that
\begin{multline}
  \label{eq:rotation}
  \ket{0}_{\mathcal{H}_m[q_h-1:1]}\ket{0}_{\mathcal{H}_m[0]}\ket{\text{pp}_{\arcsin}{\sqrt{r_h'}}}_{\mathcal{B}} \\ \mapsto \ket{0}_{\mathcal{H}_m[q_h-1:1]}\left(\sqrt{1-r_h'}\ket{0}_{\mathcal{H}_m[0]}+\sqrt{r_h'}\ket{1}_{\mathcal{H}_m[0]}\right)
  \twocolbreak
  \ket{\text{pp}_{\arcsin}{\sqrt{r_h'}}}_{\mathcal{B}}.
\end{multline}
When we write as
\begin{equation}
  \text{pp}_{\arcsin}{\sqrt{r_h'}} = \alpha_{n-1} + \alpha_{n-2}2^{-1} + \cdots + \alpha_{0}2^{1-n},
\end{equation}
Eq.~\eqref{eq:rotation} can be implemented using the quantum circuit depicted in Fig.~\ref{fig:U_sin}.
It is possible because the quantum state $\ket{h}_{\mathcal{H}_m}$ satisfies either $h\geq 2$ or $h=0$.

\begin{figure}[ht]
  \centering
  \iftwocolumnversion\resizebox{\columnwidth}{!}{\fi
  \begin{quantikz}[row sep={22pt,between origins}, column sep=8pt]
    \lstick{$\ket{h[q_{h}-1]}_{\mathcal{H}_m[q_{h}-1]}$} & \octrl{1} & \octrl{1} & \qw \ \ldots \ & \octrl{1} & \qw & \\
    & \vdots & \vdots & & \vdots \\
    \lstick{$\ket{h[1]}_{\mathcal{H}_m[1]}$} & \octrl{1}\vqw{-1} & \octrl{1}\vqw{-1} & \qw \ \ldots \ & \octrl{1}\vqw{-1} & \qw & \\
    \lstick{$\ket{h[0]}_{\mathcal{H}_m[0]}$} & \gate[1]{R_y(2)} & \gate[1]{R_y(1)} & \qw \ \ldots \ & \gate[1]{R_y(2^{2-n})} & \qw \\
    \lstick{$\ket{\alpha_{n-1}}_{\mathcal{B}[n-1]}$} & \ctrl{-1} & \qw & \qw & \qw & \qw  \\
    \lstick{$\ket{\alpha_{n-2}}_{\mathcal{B}[n-2]}$} & \qw & \ctrl{-2}  & \qw & \qw & \qw  \\
    \lstick{$\vdots$} &  &  & \ddots &  &   \\
    \lstick{$\ket{\alpha_{0}}_{\mathcal{B}[0]}$} & \qw & \qw & \qw & \ctrl{-4} & \qw
  \end{quantikz}
  \iftwocolumnversion}\fi
  \caption{\label{fig:U_sin}The quantum circuit for $U_{\sin}$.}
\end{figure}

The multicontrolled $R_y(\theta)$ gates are compiled using \cite{Nielsen2010-bp} as described in Fig.~\ref{fig:Ry}.
This circuit has 6 $R_z$ gates and an $R_z$ depth of 6.
We can reduce the $R_z$ depth by utilizing the same technique as in~\cite{Wang2024-me}.
The depth reduced circuit is depicted in Fig.~\ref{fig:Ry_reduced}.
Also, by utilizing the relationship depicted in Fig.~\ref{fig:Rz_relation}, we obtain the circuit corresponding to the multicontrolled $R_y(\theta)$ gate as depicted in Fig.~\ref{fig:Ry_final}.

\begin{figure*}
  \centering
  \resizebox{\textwidth}{!}{
  \begin{quantikz}[row sep={22pt,between origins}, column sep=8pt]
    & \octrl{1} & \qw \\
    & \vdots &    \\
    & \octrl{1} \vqw{-1} & \qw \\
    & \gate[1]{R_y(\theta)} & \qw \\
    \lstick{$\ket{0}$} & \qw & \qw \rstick{$\ket{0}$}  \\
    & \ctrl{-2} & \qw  
  \end{quantikz}
  =
  \begin{quantikz}[row sep={22pt,between origins}, column sep=2pt]
    & \octrl{1} & \qw & \qw  & \qw & \qw & \qw & \qw & \qw & \qw & \qw & \qw & \qw  & \qw & \qw & \qw & \qw & \qw & \octrl{1} & \qw \\
    & \vdots &  &  &  & & & & & & &  &  & & & & & & \vdots & \\
    & \octrl{2} \vqw{-1} & \qw & \qw  & \qw & \qw & \qw & \qw & \qw & \qw & \qw & \qw & \qw & \qw & \qw & \qw & \qw & \qw & \octrl{2} \vqw{-1} & \qw \\
    & \qw & \gate[1]{S^{\dagger}} & \gate{H} & \targ{} & \gate{R_z(-\theta/4)} & \targ{} & \gate{R_z(\theta/4)} & \targ{} & \gate{R_z(\theta/4)} & \targ{} & \gate{R_z(-\theta/4)} & \targ{} & \gate{R_z(-\theta/4)} & \targ{} & \gate{R_z(\theta/4)} & \gate{H} & \gate{S} & \qw & \qw \\
    \lstick{$\ket{0}$} & \targ{} & \qw & \qw & \qw & \qw & \qw & \ctrl{1} & \qw & \qw & \qw & \ctrl{1} & \qw & \qw & \qw & \qw & \qw & \qw & \targ{} & \qw \rstick{$\ket{0}$} \\
    & \qw & \qw & \qw & \ctrl{-2} & \qw & \ctrl{-2} & \targ{} & \ctrl{-2} & \qw & \ctrl{-2} & \targ{} & \ctrl{-2} & \qw & \ctrl{-2} & \qw & \qw & \qw & \qw & \qw
  \end{quantikz}
  }
  \caption{\label{fig:Ry}The quantum circuit for multicontrolled $R_y(\theta)$ gate.}
\end{figure*}

\begin{figure}[ht]
  \centering
  \resizebox{\columnwidth}{!}{
  \begin{quantikz}[row sep={22pt,between origins}, column sep=8pt]
    & \octrl{1} & \qw \\
    & \vdots &    \\
    & \octrl{1} \vqw{-1} & \qw \\
    & \gate[1]{R_y(\theta)} & \qw \\
    \lstick{$\ket{0}$} & \qw & \qw \rstick{$\ket{0}$}  \\
    \lstick{$\ket{0}$} & \qw & \qw \rstick{$\ket{0}$}  \\
    \lstick{$\ket{0}$} & \qw & \qw \rstick{$\ket{0}$}  \\
    \lstick{$\ket{0}$} & \qw & \qw \rstick{$\ket{0}$}  \\
    & \ctrl{-5} & \qw  
  \end{quantikz}
  =
  \begin{quantikz}[row sep={22pt,between origins}, column sep=8pt]
    & \octrl{1} & \qw & \qw  & \qw & \qw & \qw & \qw & \qw & \qw & \qw & \octrl{1} & \qw \\
    & \vdots &  &  &  & & & & & & & \vdots & \\
    & \octrl{5} \vqw{-1} & \qw & \qw  & \qw & \qw & \qw & \qw & \qw & \qw & \qw & \octrl{5} \vqw{-1} & \qw \\
    & \qw & \gate{S^{\dagger}} & \gate{H} & \ctrl{3} & \qw & \gate{R_z(\theta/4)} & \qw & \ctrl{3} & \gate{H} & \gate{S} & \qw & \qw \\
    \lstick{$\ket{0}$} & \qw & \targ{} & \qw & \targ{} & \qw & \gate{R_z(-\theta/4)} & \qw & \targ{} & \qw & \targ{} & \qw & \qw \rstick{$\ket{0}$}  \\
    \lstick{$\ket{0}$} & \qw & \qw & \qw & \targ{} & \targ{} & \gate{R_z(\theta/4)} & \targ{} & \targ{} & \qw & \qw & \qw & \qw \rstick{$\ket{0}$}  \\
    \lstick{$\ket{0}$} & \qw & \qw & \targ{} & \targ{} & \qw & \gate{R_z(-\theta/4)} & \qw & \targ{} & \targ{} & \qw & \qw & \qw \rstick{$\ket{0}$}  \\
    \lstick{$\ket{0}$} & \targ{} & \ctrl{-3} & \qw & \ctrl{1} & \qw & \qw & \qw & \ctrl{1} & \qw & \ctrl{-3} & \targ{} & \qw \rstick{$\ket{0}$}  \\
    & \qw & \qw & \ctrl{-2} & \targ{} & \ctrl{-3} & \qw  & \ctrl{-3} & \targ{} & \ctrl{-2} & \qw & \qw & \qw 
  \end{quantikz}
  }
  \caption{\label{fig:Ry_reduced}The quantum circuit for multicontrolled $R_y(\theta)$ gate with reduced depth.}
\end{figure}

\begin{figure}[htbp]
  \centering
  \resizebox{\columnwidth}{!}{
  \begin{quantikz}[row sep={22pt,between origins}, column sep=4pt]
    & \gate{R_z(\theta_1)} & \qw & \ctrl{3} & \qw & \qw & \qw & \qw & \ctrl{6} & \qw & \gate{R_z(\theta_2)} & \qw \\
    & \gate{R_z(-\theta_1)} & \qw & \targ{} & \qw & \targ{} & \qw & \qw & \qw & \qw & \qw & \qw \\
    & \gate{R_z(\theta_1)} & \targ{} & \targ{} & \qw & \qw & \qw & \qw & \qw & \qw & \qw & \qw \\
    & \gate{R_z(-\theta_1)} & \qw & \targ{} & \targ{} & \qw & \qw & \qw & \qw & \qw & \qw & \qw \\
    & \qw & \qw & \qw & \qw & \qw & \targ{} & \qw & \targ{} & \qw & \gate{R_z(-\theta_2)} & \qw \\
    & \qw & \qw & \qw & \qw & \qw & \qw & \qw & \targ{} & \targ{} & \gate{R_z(\theta_2)} & \qw \\
    & \qw & \qw & \qw & \qw & \qw & \qw & \targ{} & \targ{} & \qw & \gate{R_z(-\theta_2)} & \qw \\
    & \qw & \qw & \ctrl{1} & \qw & \ctrl{-6} & \ctrl{-3} & \qw & \ctrl{2} & \qw & \qw & \qw \\
    & \qw & \ctrl{-6} & \targ{} & \ctrl{-5} & \qw & \qw & \qw & \qw & \qw & \qw & \qw \\
    & \qw & \qw & \qw & \qw & \qw & \qw & \ctrl{-3} & \targ{} & \ctrl{-4} & \qw & \qw 
  \end{quantikz}
  =
  \begin{quantikz}[row sep={22pt,between origins}, column sep=4pt]
    & \qw & \qw & \ctrl{6} & \qw & \gate{R_z(\theta_1+\theta_2)} & \qw & \qw & \ctrl{3} & \qw & \qw & \qw \\
    & \qw & \qw & \qw & \qw & \gate{R_z(-\theta_1)} & \qw & \qw & \targ{} & \qw & \targ{} & \qw \\
    & \qw & \qw & \qw & \qw & \gate{R_z(\theta_1)} & \qw & \targ{} & \targ{} & \qw & \qw & \qw \\
    & \qw & \qw & \qw & \qw & \gate{R_z(-\theta_1)} & \qw & \qw & \targ{} & \targ{} & \qw & \qw \\
    & \targ{} & \qw & \targ{} & \qw & \gate{R_z(-\theta_2)} & \qw & \qw & \qw & \qw & \qw & \qw \\
    & \qw & \qw & \targ{} & \targ{} & \gate{R_z(\theta_2)} & \qw & \qw & \qw & \qw & \qw & \qw \\
    & \qw & \targ{} & \targ{} & \qw & \gate{R_z(-\theta_2)} & \qw & \qw & \qw & \qw & \qw & \qw \\
    & \ctrl{-3} & \qw & \ctrl{2} & \qw & \qw & \qw & \qw & \ctrl{1} & \qw & \ctrl{-6} & \qw \\
    & \qw & \qw & \qw & \qw & \qw & \qw & \ctrl{-6} & \targ{} & \ctrl{-5} & \qw & \qw \\
    & \qw & \ctrl{-3} & \targ{} & \ctrl{-4} & \qw & \qw & \qw & \qw & \qw & \qw & \qw 
  \end{quantikz}
  }
  \caption{\label{fig:Rz_relation}The relationship between the multicontrolled $R_z$ gates.}
\end{figure}

\begin{figure}[htbp]
  \centering
  \resizebox{\columnwidth}{!}{
  \begin{quantikz}[row sep={22pt,between origins}, column sep=2pt]
    \lstick{$\ket{h[q_{h}-1]}_{\mathcal{H}_m[q_{h}-1]}$} & \octrl{1} & \qw & \qw & \qw & \qw & \qw & \qw & \qw & \qw & \qw & \qw & \qw & \qw & \qw & \qw & \qw & \qw & \qw & \qw & \qw & \qw & \qw & \octrl{1} & \qw\\
    & \vdots & & & & & & & & & & & & & & & & & & & & & & \vdots & \\
    \lstick{$\ket{h[1]}_{\mathcal{H}_m[1]}$} & \octrl{12} \vqw{-1} & \qw & \qw & \qw & \qw & \qw & \qw & \qw & \qw & \qw & \qw & \qw & \qw & \qw & \qw & \qw & \qw & \qw & \qw & \qw & \qw & \qw & \octrl{12} \vqw{-1} & \qw\\
    \lstick{$\ket{h[0]}_{\mathcal{H}_m[0]}$}& \qw & \qw & \gate{S^{\dagger}} & \gate{H} & \qw & \qw & \ctrl{7} & \qw & \qw & \qw & \qw  & \gate{R_z(1-2^{1-n})} & \qw & \qw & \qw & \qw & \ctrl{7} & \qw & \qw & \gate{H} & \gate{S} & \qw & \qw & \qw \\
    \lstick{$\ket{0}$} & \qw & \targ{} & \qw & \qw & \qw & \qw & \targ{} & \qw & \qw & \qw & \qw & \gate{R_z(-2^{-1})}\gategroup[10,steps=1,style={dashed, inner sep=0pt}]{} & \qw & \qw & \qw & \qw & \targ{} & \qw & \qw & \qw & \qw & \targ{} & \qw & \qw \\
    \lstick{$\ket{0}$} & \qw & \qw & \qw & \qw & \qw & \qw & \targ{} & \targ{} & \qw & \qw & \qw & \gate{R_z(2^{-1})} & \qw & \qw & \qw & \targ{} & \targ{} & \qw & \qw & \qw & \qw & \qw & \qw & \qw \\
    \lstick{$\ket{0}$} & \qw & \qw & \targ{} & \qw & \qw & \qw & \targ{} & \qw & \qw & \qw & \qw & \gate{R_z(-2^{-1})} & \qw & \qw & \qw & \qw & \targ{} & \qw & \qw & \qw & \targ{} & \qw & \qw & \qw \\
    \lstick{$\ket{0}$} & \qw & \targ{} & \qw & \qw & \qw & \qw & \targ{} & \qw & \qw & \qw & \qw & \gate{R_z(-2^{-2})} & \qw & \qw & \qw & \qw & \targ{} & \qw & \qw & \qw & \qw & \targ{} & \qw & \qw \\
    \lstick{$\ket{0}$} & \qw & \qw & \qw & \qw & \qw & \qw & \targ{} & \qw & \targ{} & \qw & \qw & \gate{R_z(2^{-2})} & \qw & \qw & \targ{} & \qw & \targ{} & \qw & \qw & \qw & \qw & \qw & \qw & \qw \\
    \lstick{$\ket{0}$} & \qw & \qw & \qw & \targ{} & \qw & \qw & \targ{} & \qw & \qw & \qw & \qw & \gate{R_z(-2^{-2})} & \qw & \qw & \qw & \qw & \targ{} & \qw & \qw & \targ{} & \qw & \qw & \qw & \qw \\
    & & & & & \ddots & & \vdots & & & \ddots & & \vdots & & \iddots & & & \vdots & & \iddots \\
    \lstick{$\ket{0}$} & \qw & \targ{} & \qw & \qw & \qw & \qw & \targ{} & \qw & \qw & \qw & \qw & \gate{R_z(-2^{-n})} & \qw & \qw & \qw & \qw & \targ{} & \qw & \qw & \qw & \qw & \targ{} & \qw & \qw \\
    \lstick{$\ket{0}$} & \qw & \qw & \qw & \qw & \qw & \qw & \targ{} & \qw & \qw & \qw & \targ{} & \gate{R_z(2^{-n})} & \targ{} & \qw & \qw & \qw & \targ{} & \qw & \qw & \qw & \qw & \qw & \qw & \qw \\
    \lstick{$\ket{0}$} & \qw & \qw & \qw & \qw & \qw & \targ{} & \targ{}\vqw{-3} & \qw & \qw & \qw & \qw & \gate{R_z(-2^{-n})} & \qw & \qw & \qw & \qw & \targ{}\vqw{-3} & \targ{} & \qw & \qw & \qw & \qw & \qw & \qw \\
    \lstick{$\ket{0}$} & \targ{} & \ctrl{-10} & \qw & \qw & \qw & \qw & \ctrl{3} & \qw & \qw & \qw & \qw & \qw & \qw & \qw & \qw & \qw & \ctrl{3} & \qw & \qw & \qw & \qw & \ctrl{-10} & \targ{} & \qw \\
    \lstick{$\ket{\alpha_{n-1}}_{\mathcal{B}[n-1]}$} & \qw & \qw & \ctrl{-9} & \qw & \qw & \qw & \targ{} & \ctrl{-10} & \qw & \qw & \qw & \qw & \qw & \qw & \qw & \ctrl{-10} & \targ{} & \qw & \qw & \qw & \ctrl{-9} & \qw & \qw & \qw \\
    \lstick{$\ket{\alpha_{n-2}}_{\mathcal{B}[n-2]}$} & \qw & \qw & \qw & \ctrl{-7} & \qw & \qw & \targ{} & \qw & \ctrl{-8} & \qw & \qw & \qw & \qw & \qw & \ctrl{-8} & \qw & \targ{} & \qw & \qw & \ctrl{-7} & \qw & \qw & \qw & \qw \\
    & & & & & \ddots & & \vdots & & & \ddots & & & & \iddots & & & \vdots & & \iddots \\
    \lstick{$\ket{\alpha_{0}}_{\mathcal{B}[0]}$} & \qw & \qw & \qw & \qw & \qw & \ctrl{-5} & \targ{}\vqw{-1} & \qw & \qw & \qw & \ctrl{-6} & \qw & \ctrl{-6} & \qw & \qw & \qw & \targ{}\vqw{-1} & \ctrl{-5} & \qw & \qw & \qw & \qw & \qw & \qw 
  \end{quantikz}
  }
  \caption{\label{fig:Ry_final}The quantum circuit for multicontrolled $R_y(\theta)$ gate.}
\end{figure}

The area enclosed by the dashed line in Fig.~\ref{fig:Ry_final} can be implemented effectively by utilizing the recursive phase catalysis circuit described in~ \cite{Wang2024-me}.
The total cost required for $U_{\sin}$ is:
\begin{align}
  T_\text{count}(U_{\sin}) &= 12n_{\epsilon}+6.6\log_2(4/\epsilon_{\sin}) + 8q_h - 16, \\
  T_\text{depth}(U_{\sin}) &= 3n_{\epsilon}+1.15\log_2(4/\epsilon_{\sin})+ 2q_h - 3, \\
  a_\text{count}(U_{\sin}) &= 5n_{\epsilon}+2,
\end{align}
where $T_\text{depth}(*)$ and $a_\text{count}(*)$ denote the T-depth and the number of auxiliary qubits of the operation $*$, respectively, $n_{\epsilon}$ is the number of qubits of $\mathcal{B}$, $q_h$ is the number of qubits of $\mathcal{H}_m$, and $\epsilon_{\sin}$ is the error tolerance of the controlled rotations in this operation.

\subsection{Quantum algorithm for $U_Q$}
The $U_Q$ gate uncomputes the auxiliary states stored in the qubits $\mathcal{B}$ and $\mathcal{C}$ while updating $s_{h+1}$ to $s_{h}$.
The quantum circuit for $U_Q$ is composed of the following operations:
\begin{enumerate}
  \item Conduct inverse operation of steps 10--4 in $U_P$ gate and uncompute the auxiliary qubits $\mathcal{B}, \mathcal{C}_5$ to $\mathcal{C}_9$:
  \begin{multline}
    \ket{s_{h+1}}_{\mathcal{A}}\ket{\text{pp}_{\arcsin}{\sqrt{r_h'}}}_{\mathcal{B}}\ket{r_{h}}_{\mathcal{C}_4}\ket{z}_{\mathcal{C}_5}\ket{w}_{\mathcal{C}_6}
    \twocolbreak
    \ket{\sqrt{w}}_{\mathcal{C}_7}\ket{\sqrt{s_{h+1}}}_{\mathcal{C}_8}\ket{\sqrt{w/s_{h+1}}}_{\mathcal{C}_9} \\ \mapsto \ket{s_{h+1}}_{\mathcal{A}}\ket{0}_{\mathcal{B}}\ket{r_{h}}_{\mathcal{C}_4}\ket{0}_{\mathcal{C}_5}\ket{0}_{\mathcal{C}_6}\twocolbreak\ket{0}_{\mathcal{C}_7}\ket{0}_{\mathcal{C}_8}\ket{0}_{\mathcal{C}_9}.
  \end{multline}
  \item Compute the $s_{h}=s_{h+1}-r_h$ from $r_{h}$ and $s_{h+1}$: 
  \begin{equation}
    \ket{s_{h+1}}_{\mathcal{A}}\ket{r_{h}}_{\mathcal{C}_4} \mapsto \ket{s_{h}}_{\mathcal{A}}\ket{r_{h}}_{\mathcal{C}_4}.
  \end{equation}
  \item Conduct inverse operation of steps 3, 2, and 1 in $U_P$ gate and uncompute the auxiliary qubits $\mathcal{C}_1$ to $\mathcal{C}_4$:
  \begin{multline}
    \ket{n_{i(h)}}_{\mathcal{N}_{i(h)}}\ket{n_{j(h)}}_{\mathcal{N}_{j(h)}}\ket{n_{i(h)}}_{\mathcal{C}_1}\ket{n_{j(h)}}_{\mathcal{C}_2}\twocolbreak\ket{n_{i(h)}n_{j(h)}}_{\mathcal{C}_3}\ket{r_{h}}_{\mathcal{C}_4} \\ \mapsto \ket{n_{i(h)}}_{\mathcal{N}_{i(h)}}\ket{n_{j(h)}}_{\mathcal{N}_{j(h)}}\twocolbreak\ket{0}_{\mathcal{C}_1}\ket{0}_{\mathcal{C}_2}\ket{0}_{\mathcal{C}_3}\ket{0}_{\mathcal{C}_4}.
  \end{multline}
\end{enumerate}
In total, the $U_Q$ gate requires 1 $\text{MUL\_INT}_{q_1,q_1}$, 1 $\text{MUL\_CONST\_INT\_UI}_{2q_1,n_{\epsilon}}$, 1 $\text{COMP}_{n_{\epsilon}}$, 2 $\text{c-SUB}_{n_{\epsilon}}$, 2 $\text{SQRT}_{n_{\epsilon}}$, 1 $\text{DIV}_{n_{\epsilon}}$, 1 $\text{ARCSIN}_{n_{\epsilon},\epsilon_{\arcsin}}$, and 1 $\text{SUB}_{n_{\epsilon}}$.

\subsection{Quantum algorithm for $U_{\text{add}}$}
The $U_{\text{add}}$ gate adds constant 1 to the history register when the history register is not $\ket{0}_{\mathcal{H}_m}$ and does nothing when the history register is $\ket{0}_{\mathcal{H}_m}$.
The quantum circuit for $U_{\text{add}}$ is composed of the following operations:
\begin{enumerate}
  \item Add 1 to the history register $\mathcal{H}_m$:
  \begin{equation}
    \sum_{k=0}^{h-1}\ket{k}_{\mathcal{H}_m} \mapsto \sum_{k=0}^{h-1}\ket{k+1}_{\mathcal{H}_m}=\sum_{k=1}^{h}\ket{k}_{\mathcal{H}_m}.
  \end{equation}
  In this step, $\text{ADD\_CONST}_{q_h}$ is used.
  \item Apply CNOT gate to the least significant digit of the history register $\mathcal{H}_m$ to restore to its original zero state:
  \begin{equation}
    \sum_{k=1}^{h}\ket{k}_{\mathcal{H}_m} \mapsto \ket{0}_{\mathcal{H}_m} + \sum_{k=2}^{h}\ket{k}_{\mathcal{H}_m}.
  \end{equation}
  In this step, $\text{Toffoli}_{q_h}$ is used.
\end{enumerate}
In total, the $U_{\text{add}}$ gate requires 1 $\text{ADD\_CONST}_{q_h}$ and 1 $\text{Toffoli}_{q_h}$.

\subsection{Quantum algorithm for $U_R$}
The $U_R$ gate uncomputes the auxiliary states stored in the qubits $\mathcal{A}$ as 
\begin{equation}
  \ket{s_{h}}_{\mathcal{A}} \mapsto \ket{s_{h+1}}_{\mathcal{A}}.
\end{equation}
The quantum circuit for $U_R$ is composed of the inverse operation of steps 2 and 3 of $U_Q$ and the inverse operation of steps 1--3 of $U_P$ gate.
The total cost required for $U_R$ is 2 $\text{MUL\_INT}_{q_1,q_1}$, 2 $\text{MUL\_CONST\_INT\_UI}_{2q_1,n_{\epsilon}}$, and 1 $\text{ADD}_{n_{\epsilon}}$.

\subsection{Quantum algorithm for $U_{\text{shift}}$}
The $U_{\text{shift}}$ gate shifts the mass distribution $\bar{\bm{n}}$ to $\bar{\bm{n}}'$ according to the transition label $h_m$ stored in the history register $\mathcal{H}_m$.
The process is one of the following:
\begin{enumerate}
  \item The number of droplets in the $i$-th and $j$-th bins decreases by one, and the number of droplets in the $i+j$-th bin increases by one.
  \item The number of droplets in the $i$-th bin decreases by two, and the number of droplets in the $2i$-th bin increases by one.
\end{enumerate}
In either case, the shift can be realized by utilizing multi-controlled ADD (SUB) gate.
The multicontrolled ADD (SUB) gate can be organized by utilizing Toffoli gate and c-ADD (c-SUB) gate.
Therefore, the total cost required for the $U_{\text{shift}}$ gate is 2 $\text{Toffoli}_{q_h}$, 1 $\text{c-ADD}_{q_{i+j}}$, 1 $\text{c-SUB}_{q_i}$, and 1 $\text{c-SUB}_{q_j}$ for the first case and 2 $\text{Toffoli}_{q_h}$, 1 $\text{c-ADD}_{q_{2i}}$, and 1 $\text{c-SUB}_{q_i}$ for the second case.

\subsection{Quantum algorithm for $U_c$}
The $U_c$ gate is composed of repeated applications of $R_y$ gates.
The number of repetitions is determined by $n_i^{\max}$ for the $i$-th bin.
The total cost required for $U_c$ is
\begin{equation}
  T_\text{count}(U_c) = 1.15n_i^{\max}\log_2(n_i^{\max}/\epsilon_c),
\end{equation}
where $n_i^{\max}=\lfloor N/i \rfloor$ is the maximum number of droplets in the $i$-th bin and $\epsilon_c$ is the error tolerance for this operation.
We utilize the repeat until success method~\cite{Bocharov2015-km} to implement the $R_y$ gates.

\bibliography{bib}

\end{document}